\documentclass{llncs}

\usepackage{xspace,amsmath,url}
\usepackage{arydshln}

\usepackage{graphics}
\usepackage{epstopdf}
\usepackage{epsfig}

\newcommand{\exclude}[1]{}

\begin{document}

\title{Sampling the suffix array with minimizers}

\author{Szymon Grabowski
and
Marcin Raniszewski
}

\institute{
	  Lodz University of Technology, Institute of Applied Computer Science,\\
	  Al.\ Politechniki 11, 90--924 {\L}\'od\'z, Poland
	  \email{\{sgrabow|mranisz\}@kis.p.lodz.pl}
}

\maketitle

\begin{abstract}
Sampling (evenly) the suffixes from the suffix array is an old idea trading 
the pattern search time for reduced index space. 
A few years ago Claude et al. showed an alphabet sampling scheme allowing 
for more efficient pattern searches compared to the sparse 
suffix array, for long enough patterns.
A drawback of their approach is the requirement that sought patterns 
need to contain at least one character from the chosen subalphabet.
In this work we propose an alternative suffix sampling approach 
with only a minimum pattern length as a requirement, which seems
more convenient in practice.
Experiments show that our algorithm achieves competitive time-space tradeoffs 
on most standard benchmark data.
\end{abstract}

\section{Introduction}
\noindent 
Full-text indexes built over a text of length $n$ 
can roughly be divided into two categories:
those requiring at least $n\log_2 n$ bits and the more compact ones.
Classical representatives of the first group are the suffix tree 
and the suffix array.
Succinct solutions, often employing the Burrows--Wheeler transform 
and other ingenious mechanisms (compressed rank/select data structures, 
wavelet trees, etc.), are object of vivid interest in theoretical 
computer science~\cite{NM2007}, but their practical performance 
does not quite deliver; 
in particular, the locate query is significantly slower than using 
e.g. the suffix array~\cite{PST2006,FGNV2008,GP2013}.

A very simple, yet rather practical alternative to both compressed 
indexes and the standard suffix array is the {\em sparse suffix array} 
(SpaSA)~\cite{DBLP:conf/cocoon/KarkkainenU96}.
This data structure stores only the suffixes at regular positions, 
namely those being a multiple of $q$ 
($q > 1$ is a construction-time parameter).
The main drawback of SpaSA is that instead of one (binary) search over 
the plain SA it has to perform $q$ searches, in $q-1$ cases of which 
followed by verification of the omitted prefix against the text.
If, for example, the pattern $P[1 \ldots 6]$ is \texttt{tomcat} and $q = 4$, 
we need to search for \texttt{tomcat}, \texttt{omcat}, \texttt{mcat} 
and \texttt{cat}, and 3 of these 4 searches will be followed by verification.
Obviously, the pattern length must be at least $q$ and this approach
generally works better for longer patterns.

The {\em sampled suffix array} (SamSA) by Claude et al.~\cite{CNPSTjda10}
is an ingenious alternative to SpaSA.
They choose a subset of the alphabet and build a sorted array over only 
those suffixes which start with a symbol from the chosen subalphabet.
The search starts with finding the first (leftmost) sampled 
symbol of the pattern, let us say at position $j$, 
and then the pattern suffix $P[j \ldots m]$ is sought in the sampled suffix 
array with standard means.
After that, each occurrence of the pattern suffix must be verified 
in the text with the previous $j-1$ symbols.
A great advantage of SamSA over SpaSA is that it performs only 
one binary search.
On the other hand, a problem is that the pattern must contain at least 
one symbol from the sampled subalphabet.
It was shown however that a careful selection of the subalphabet 
allows for leaving out over 80\% suffixes and still almost preserving
the pattern search speed for the standard array, 
if the patterns are long (50--100).

An idea most similar to ours was presented more than a decade ago 
by Crescenzi et al.~\cite{CLGLPR00,CLGLPR03} and was called 
text sparsification via local maxima.
Using local maxima, that is, symbols in text which are lexicographically 
not smaller than the symbol just before them and lexicographically greater 
than the next symbol, has been recognized even earlier as a useful technique 
in string matching and dynamic data structures, for problems like 
indexing dynamic texts~\cite{ABR00},
maintaining dynamic sequences under equality tests~\cite{MSU97} 
or parallel construction of suffix trees~\cite{SV94}.
Crescenzi et al., like us, build a suffix array on sampled suffixes, 
yet in their experiments (only on DNA) the index compression by factor 
about 3 requires patterns of length at least about 150 
(otherwise at least a small number of matches are lost).
Our solution does not suffer a similar limitation, that is, 
the minimum pattern lengths with practical parameter settings 
are much smaller.

\section{Our algorithm}
\label{sec:alg1}

\subsection{The idea}

Our purpose is to combine the benefits of the sparse suffix array 
(searching any patterns of length at least the sampling parameter $q$) 
and the sampled suffix array (one binary search).
To this end, we need the following property:

{\em For each substring $s$ of $T$, $|s| = q$, there exists its 
substring $s'$, $|s'| \leq q$, such that among the sampled suffixes 
there exists at least one which starts with $s'$.
Moreover, $s'$ is obtained from $s$ deterministically, or in other words: 
for any two substrings of $T$, $s_1$ and $s_2$, if $s_1 = s_2$, 
then $s'_1 = s'_2$.}

This property is satisfied if a {\em minimizer} of $s$ is taken as $s'$.
The idea of minimizers was proposed by Roberts et al. in 2004~\cite{RHHMY2004} 
and seemingly first overlooked in the bioinformatics (or string matching) 
community, only to be revived in the last years 
(cf., e.g.,~\cite{MFC2012,LKHYYS2013,CLJSM2014,WS2014}). 
The minimizer for a sequence $s$ of length $r$ is the lexicographically 
smallest of its all $(r-p+1)$ $p$-grams (or $p$-mers, in the term commonly 
used in computational biology); usually it is assumed that $p \ll r$.
For a simple example, note that two DNA sequencing reads 
with a large overlap are likely to share the same minimizer, 
so they can be clustered together.
That is, the smallest $p$-mer may be the identifier of the bucket into 
which the read is then dispatched.

Coming back to our algorithm: in the construction phase, we pass a sliding 
window of length $q$ over $T$ and calculate the lexicographically smallest 
substring of length $p$ in each window (i.e., its minimizer).
Ties are resolved in favor of the leftmost of the smallest substrings.
The positions of minimizers are start positions of the sampled suffixes, 
which are then lexicographically sorted, like for a standard suffix array.
The values of $q$ and $p$, $p \leq q$, are construction-time parameters.

In the actual construction, we build a standard suffix array 
and in extra pass over the sorted suffix indexes copy the sampled ones into 
a new array.
This requires an extra bit array of size $n$ for storing the sampled suffixes 
and in total may take $O(n)$ time and $O(n)$ words of space.
In theory, we can use one of two randomized algorithms by I et al.~\cite{IKK2014} 
which sort $n' = o(n)$ arbitrary suffixes of text of length $n$ 
either in $O(n)$ time using $O(n'\log n')$ words of space (Monte Carlo 
algorithm), or in $O(n \log n')$ time using $O(n')$ words of space 
(Las Vegas algorithm).

There is a small caveat: 
the minimizer at the new sampled position may be equal to the previous one, 
if only the window has been shifted beyond the position of its previous 
occurrence.
The example illustrates.
We set $q = 5$, $p = 1$ and the text is $T = Once\_upon\_a\_time$.
In the first window ($Once\_$) the minimizer will be the blank space 
and it does not change until $\_upon$ (including it), 
but the next window ($upon\_$) also has a blank space as its minimizer, 
yet it is a new string, in a different position.
Both blank spaces are thus prefixes of the suffixes to sample.

The search is simple: in the prefix $P[1 \ldots q]$ of the pattern 
its minimizer is first found, at some position $1 \leq j \leq q - p + 1$, 
and then we binary search the pattern suffix $P[j \ldots m]$, 
verifying each tentative match with its truncated $(j-1)$-symbol prefix 
in the text.

Note that any other pattern window 
$P[i \ldots i+q-1]$, $2 \leq i \leq m-q+1$, 
could be chosen to find its minimizer and continue the search over 
the sampled suffix array, but using no such window can result in a 
narrower range of suffixes to verify than the one obtained from 
the pattern prefix.
This is because for any non-empty string $s$ with 
$occ_s$ occurrences in text $T$, we have $occ_s \geq occ_{xs}$, where 
$xs$ is the concatenation of a non-empty string $x$ and string $s$.

We call the described algorithm as the {\em sampled suffix array 
with minimizers} (SamSAMi).

\subsection{Parameter selection}

There are two free parameters in SamSAMi, the window length $q$ 
and the minimizer length $p$, $p \leq q$.
Naturally, the case of $p = q$ is trivial (all suffixes sampled, 
i.e. the standard suffix array obtained).
For a settled $p$ choosing a larger $q$ has a major benefit:
the expected number of selected suffixes diminishes, which reduces 
the space for the structure.
On the other hand, it has two disadvantages: $q$ is also the minimum 
pattern length, which excludes searches for shorter patterns, 
and for a given pattern length $m \geq q$ the average length of its 
sought suffix $P[j \ldots m]$ decreases, which implies more 
occcurrence verifications.

For a settled $q$ the optimal choice of the minimizer length $p$ is 
not easy; too small value (e.g., 1) may result in frequent changes 
of the minimizer, especially for a small alphabet, but 
on the other hand its too large value has the same effect, 
since a minimizer can be unchanged over at most
$p - q + 1$ successive windows.
Yet, the pattern suffix to be sought has in the worst case 
exactly $p$ symbols, which may be a suggestion that $p$ should not be 
very small.

\subsection{Speeding up the verifications}
\label{sec:alg1extrabits}

For some texts and large value of $q$ the number of verifications on 
the pattern prefix symbols tends to be large. 
Worse, each such verification requires a lookup to the text with a likely cache miss.
We propose a simple idea reducing the references to the text.

To this end, we add an extra 4 bits to each SamSAMi offset. 
Their store the distance to the previous sampled minimizer in the text.
In other words, the list of distances corresponds to the differences between 
successive SamSAMi offsets in text order.
For the first sampled minimizer in the text and any case where the difference exceeds 15 
(i.e., could not fit 4 bits), we use the value 0.
To give an example, if the sampled text positions are: 3, 10, 12, 15, 20, 
then the list of differences is: 0, 7, 2, 3, 5.
In our application the extra 4 bits are kept in the 4 most significant bits of the offset, 
which restricts the 32-bit offset usage to texts up to 256\,MB.

In the search phase, we start with finding the minimizer for $P[1 \ldots q]$, 
at some position $1 \leq \ell \leq q-p+1$,
and for each corresponding suffix from the index we read the distance to the previous minimizer 
in the text.
If its position is aligned in front of the pattern, or the read 4 bits hold the value 0,
we cannot draw any conclusion and follow with a standard verification.
If however the previous minimizer falls into the area of the (aligned) pattern, 
in some cases we can conclude that the previous $\ell-1$ symbols from the text 
do not match the $\ell-1$ long prefix of the pattern.
Let us present an example.
$P = \texttt{ctgccact}$, $q = 5$, $p = 2$.
The minimizer in the $q$ long prefix of $P$ is \texttt{cc}, starting at position $P[4]$.
Assume that $P$ is aligned with a match in $T$.
If we shift the text window left by 1 symbol and consider its minimizer, 
it may be either \texttt{ct} (corresponding to $P[1 \dots 2]$), or \texttt{?c}, 
where \texttt{?} is an unknown symbol aligned just before the 
pattern and \texttt{c} aligned with $P[1]$, if \texttt{?c} happens to be 
lexicographically smaller than \texttt{ct}.
If, however, the distance written on the 4 bits associated with the 
suffix \texttt{ccact...} of $T$ is 1 or 2, we know that we have a mismatch 
on the pattern prefix and the verification is over (without looking up the text), 
since neither \texttt{gc} or \texttt{tg} cannot be the previous minimizer.
Finally, if the read value is either 0 or at least 4, we cannot make use of this 
information and invoke a standard verification.

\subsection{SamSAMi-hash}
\label{sec:alg2}

In~\cite{GR2014} we showed how to augment the standard suffix array with 
a hash table, to start the binary search from a much more narrow interval. 
The start and end position in the suffix array for each range of suffixes 
having a common prefix of length $k$ was inserted into the hash table, 
where the key for which the hash function was calculated was the prefix string.
The same function was applied to the pattern's prefix and after a HT lookup 
the binary search was continued with reduced number of steps.
The mechanism requires $m \geq k$.
To estimate the space needed by the extra table, the reader is advised 
to look at Table~\ref{table:qgrams}, presenting the number of distinct 
$q$-grams in five 200\,MB datasets from the popular Pizza~\&~Chili 
text corpus.
For example for the text \texttt{english} the number of distinct 8-grams 
is 20,782,043, which is about 10\% of the text length.
This needed to be multiplied by 16 in our implementation (open addressing with 
linear probing and 50\% load factor and two 4-byte integers per entry), 
which results in about 1.6$n$ bytes overhead.

\begin{table}
\centering
\begin{tabular}{lrrrrr}
\hline
~~$q$   &~~~~~~~~~~~~~~~\texttt{dna}~~~&~~~~~~\texttt{english}~~~&~~~~\texttt{proteins}~~~&~~~~~\texttt{sources}~~~&~~~~~~~~~~~~~\texttt{xml}~~~\\
\hline
~~1 & 16~~~& 225~~~& 25~~~& 230~~~& 96~~~\\
~~2 & 152~~~& 10,829~~~& 607~~~& 9,525~~~& 7,054~~~\\
~~3 & 683~~~& 102,666~~~& 11,607~~~& 253,831~~~& 141,783~~~\\
~~4 & 2,222~~~& 589,230~~~& 224,132~~~& 1,719,387~~~& 908,131~~~\\
~~5 & 5,892~~~& 2,150,525~~~& 3,623,281~~~& 5,252,826~~~& 2,716,438~~~\\
~~6 & 12,804~~~& 5,566,993~~~& 36,525,895~~~& 10,669,627~~~& 5,555,190~~~\\
~~7 & 28,473~~~& 11,599,445~~~& 94,488,651~~~& 17,826,241~~~& 8,957,209~~~\\
~~8 & 80,397~~~& 20,782,043~~~& 112,880,347~~~& 26,325,724~~~& 12,534,152~~~\\
\hline
\end{tabular}
\vspace{4mm}
\caption{The number of distinct $q$-grams ($1 \ldots 8$) in the datasets.
Each dataset is of length 209,715,200 bytes.}
\label{table:qgrams}
\end{table}

We can adapt this idea to SamSAMi. 
Again, the hashed keys will be $k$-long prefixes, yet now each of the sampled 
suffixes starts with some minimizer (or its prefix). 
We can thus expect a smaller overhead.
Its exact value for a particular dataset depends on three parameters, 
$k$, $q$ and $p$.
Note however that now the pattern length $m$ must be at least 
$\max(q - p + k, q)$.

\subsection{Compressing the text}
\label{sec:text_compr}

All SA-like indexes refer to the text, so to reduce the overall space 
we can compress it.
It is possible to apply a standard solution to it, like Huffman or 
Hu--Tucker~\cite{HT1971} coding (where the idea of the latter is to 
preserve lexicographical order 
of the code and thus enable direct string comparisons between the compressed 
pattern and the compressed text), 
but in SamSAMi it is more convenient to compress the text with aid of minimizers.
More precisely, we partition $T[1 \ldots n]$ into phrases: 
$T[1 \ldots j_1], T[j_1+1 \ldots j_2], \ldots, T[j_{n'-1}+1 \ldots n]$, 
$n' \leq n$, $j_1 \geq 0$,
where each 
$T[j_i + 1]$ location is a start position of a new minimizer, considering all 
$q$-long text windows moved from the beginning to the end of the text,
for the chosen parameters $q$ and $p$.
Note that $n'/n$ is the compression ratio (between 0 and 1) of the suffix array 
sampling.
The resulting sequence of phrases $T'[1 \ldots n']$ is then compressed 
with a byte code~\cite{BFNP2007}.
The dictionary of phrases $\mathcal{D}$ has to be stored too.
We note that $q$ shouldn't be too large in this variant, 
otherwise the phrases will tend to have a single occurrence and the 
dictionary representation will be dominating in space use.

In this variant we assume that $m \geq 2q - p + 1$. 
Searching for the pattern proceeds as follows.
First the minimizer in $P[1 \ldots q]$ is found, 
at some position $1 \leq j_1 \leq q - p + 1$.
Then the minimizer in $P[j_1+1 \ldots j_1+q]$ is found, 
at some position $j_1+1 \leq j_2 \leq j_2+q-p+1$.
This means that the pattern comprises the phrase $P[j_1 \ldots j_2-1]$.
This phrase is encoded with its codeword in $\mathcal{D}$.
If $P[j_2 \ldots m]$ comprises $k$ extra phrases, $k \geq 1$, 
then all of them are also translated to their codewords from $\mathcal{D}$.
The resulting concatenation of codewords for $k+1$ phrases, 
spanning $P[j_1 \ldots j_{k+1}-1]$ in the pattern, 
is the artificial pattern $P'$ to be binary searched in the suffix array 
with $n'$ sampled suffixes.
Still, all the suffixes in the range starting with the encoding of $P'$ 
have to be verified, both with the pattern prefix (of length $j_1 - 1$) 
and pattern suffix (of length $m - j_{k+1} + 1$).
Each candidate occurrence is verified with decoding its preceding phrase 
in the text and then performing a comparison on the prefix, 
and decoding its following phrase in text with an analogous comparison.

We note that the same text encoding can be used for online pattern search 
(cf.~\cite{FG2006}).

\section{Experimental results}
\label{sec:exp}

We have implemented three variants of the SamSAMi index: 
the basic one (denoted as \texttt{SamSAMi} on the plots), 
the one with reduced verifications (\texttt{SamSAMi2}) 
and the basic one augmented with a hash table (\texttt{SamSAMi-hash}).
We compared them against the sparse suffix array (\texttt{SpaSA}), 
in our implementation,
and the sampled suffix array (\texttt{SamSA})~\cite{CNPSTjda10}, 
using the code provided by its authors.
Compression of the text (Sect.~\ref{sec:text_compr}) has not been implemented.

All experiments were run on a computer with an Intel i7-4930K 3.4\,GHz CPU,
equipped with 64\,GB of DDR3 RAM and running Ubuntu 14.04 LTS 64-bit.
All codes were written in C++ and compiled with g++ 4.8.2 with -03 option.

\begin{table}
\centering
\begin{tabular}{rrrrrrr}
\hline
~~~~$q$~~&~~$p$~~&~~~~~~~~~\texttt{dna}~~&~~~\texttt{english}~~&~~\texttt{proteins}~~&~~~~\texttt{sources}~~&~~~~~~~~~~\texttt{xml}~~\\
\hline
 ~~4~~&~~1~~& 46.1~~& 39.7~~& 40.5~~& 46.1~~& 45.8~~\\
 ~~4~~&~~2~~& 55.2~~& 51.0~~& 51.0~~& 55.8~~& 54.1~~\\[2ex]
 ~~5~~&~~1~~& 40.9~~& 32.3~~& 34.0~~& 38.8~~& 39.3~~\\
 ~~5~~&~~2~~& 44.9~~& 39.9~~& 40.8~~& 46.2~~& 45.9~~\\[2ex]
 ~~6~~&~~1~~& 37.6~~& 27.7~~& 29.4~~& 34.5~~& 32.5~~\\
 ~~6~~&~~2~~& 38.0~~& 32.3~~& 34.1~~& 38.8~~& 39.3~~\\[2ex]
 ~~8~~&~~1~~& 33.7~~& 22.1~~& 23.2~~& 28.3~~& 22.0~~\\
 ~~8~~&~~2~~& 29.5~~& 23.8~~& 25.5~~& 30.5~~& 26.6~~\\[2ex]
 ~~10~~&~~1~~& 31.8~~& 19.3~~& 19.4~~& 25.0~~& 17.1~~\\
 ~~10~~&~~2~~& 24.5~~& 18.5~~& 20.5~~& 25.9~~& 18.5~~\\
 ~~10~~&~~3~~& 25.8~~& 20.8~~& 22.7~~& 27.9~~& 21.9~~\\[2ex]
 ~~12~~&~~1~~& 30.7~~& 17.9~~& 16.8~~& 22.5~~& 13.7~~\\
 ~~12~~&~~2~~& 21.2~~& 15.4~~& 17.1~~& 22.8~~& 15.1~~\\
 ~~12~~&~~3~~& 21.4~~& 16.8~~& 18.6~~& 24.2~~& 17.0~~\\[2ex]
 ~~16~~&~~1~~& 29.7~~& 16.4~~& 13.7~~& 19.3~~& 11.0~~\\
 ~~16~~&~~2~~& 17.1~~& 12.0~~& 12.9~~& 18.6~~& 11.3~~\\
 ~~16~~&~~3~~& 16.1~~& 12.6~~& 13.7~~& 19.4~~& 11.9~~\\[2ex]
 ~~24~~&~~2~~& 13.3~~& 8.4~~& 8.7~~& 13.6~~& 7.1~~\\
 ~~24~~&~~3~~& 11.1~~& 8.7~~& 9.0~~& 13.9~~& 7.4~~\\[2ex]
 ~~32~~&~~2~~& 11.7~~& 6.5~~& 6.6~~& 10.6~~& 5.1~~\\
 ~~32~~&~~3~~& 8.7~~& 6.7~~& 6.7~~& 10.6~~& 5.4~~\\[2ex]
 ~~40~~&~~2~~& 10.8~~& 5.3~~& 5.3~~& 8.5~~& 4.2~~\\
 ~~40~~&~~3~~& 7.3~~& 5.4~~& 5.3~~& 8.4~~& 4.3~~\\[2ex]
 ~~64~~&~~2~~& 9.8~~& 2.9~~& 3.4~~& 4.7~~& 3.1~~\\
 ~~64~~&~~3~~& 5.4~~& 3.0~~& 3.3~~& 4.4~~& 2.6~~\\
 ~~64~~&~~4~~& 4.4~~& 3.1~~& 3.4~~& 4.3~~& 2.7~~\\[2ex]
 ~~80~~&~~2~~& 9.6~~& 1.9~~& 2.7~~& 3.5~~& 2.9~~\\
 ~~80~~&~~3~~& 4.8~~& 1.8~~& 2.7~~& 3.1~~& 2.2~~\\
 ~~80~~&~~4~~& 3.7~~& 1.9~~& 2.7~~& 3.0~~& 2.2~~\\[2ex]
\hline
\end{tabular}
\vspace{4mm}
\caption{The percentage of suffixes that are sampled using 
the idea of minimizers with the parameters $q$ and $p$}
\label{table:pq}
\end{table}

We start with finding the fraction of sampled suffixes for multiple 
$(q, p)$ parameter pairs and the five 50\,MB Pizza~\&~Chili datasets.
Table~\ref{table:pq} presents the results.

\begin{figure}[pt]
\centerline{
\includegraphics[width=0.49\textwidth,scale=1.0]{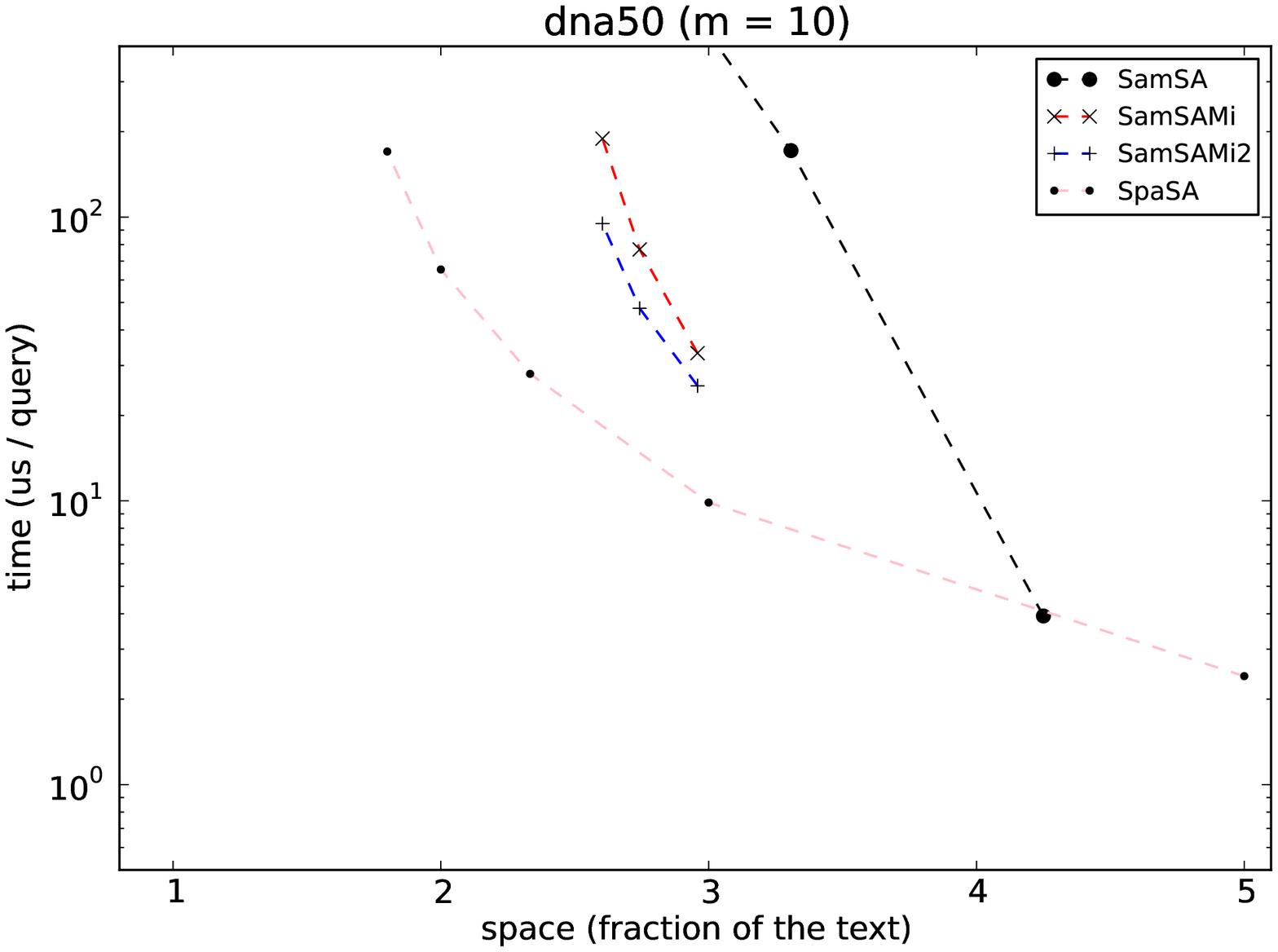}
\includegraphics[width=0.49\textwidth,scale=1.0]{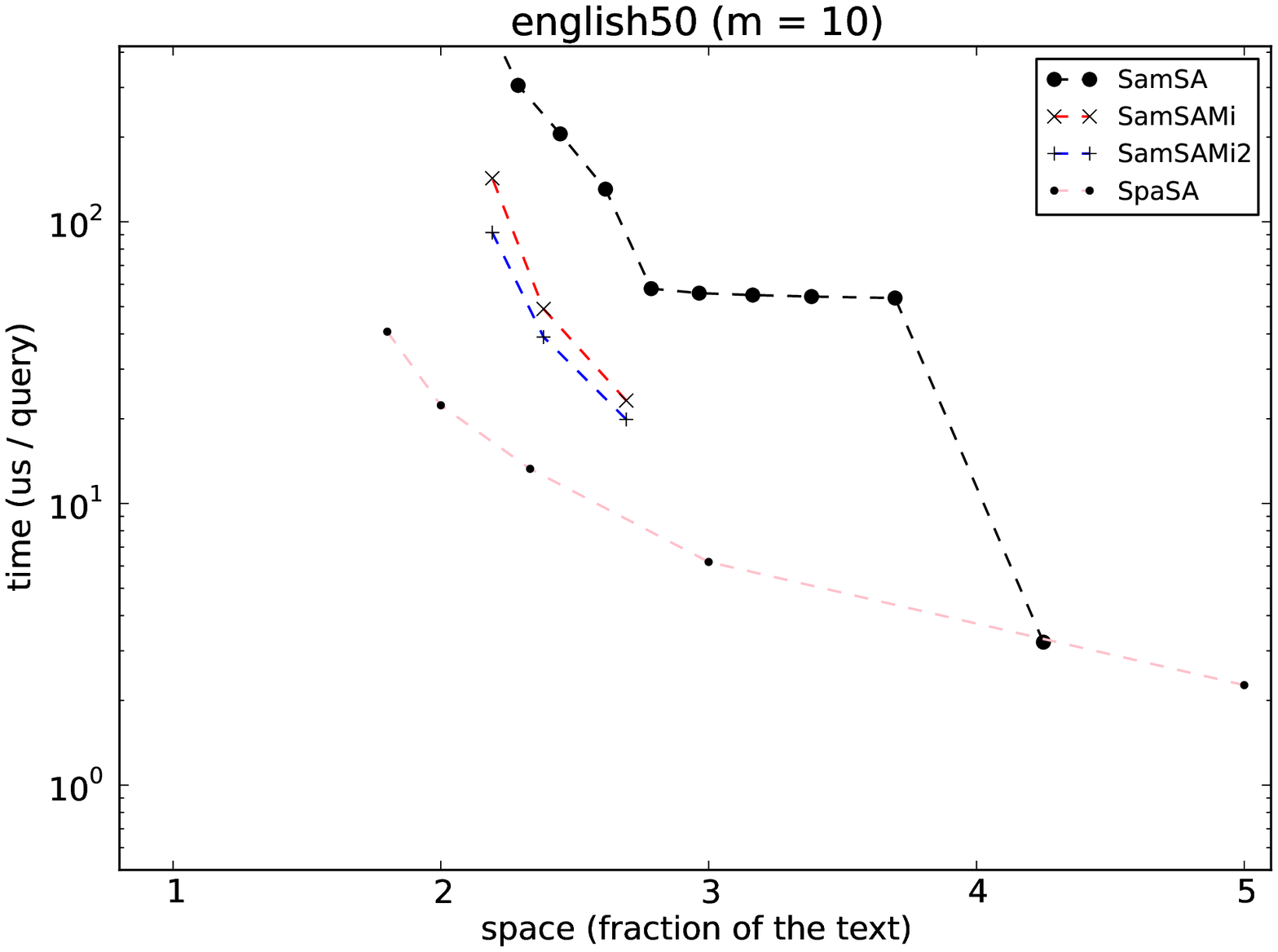}
}
\centerline{
\includegraphics[width=0.49\textwidth,scale=1.0]{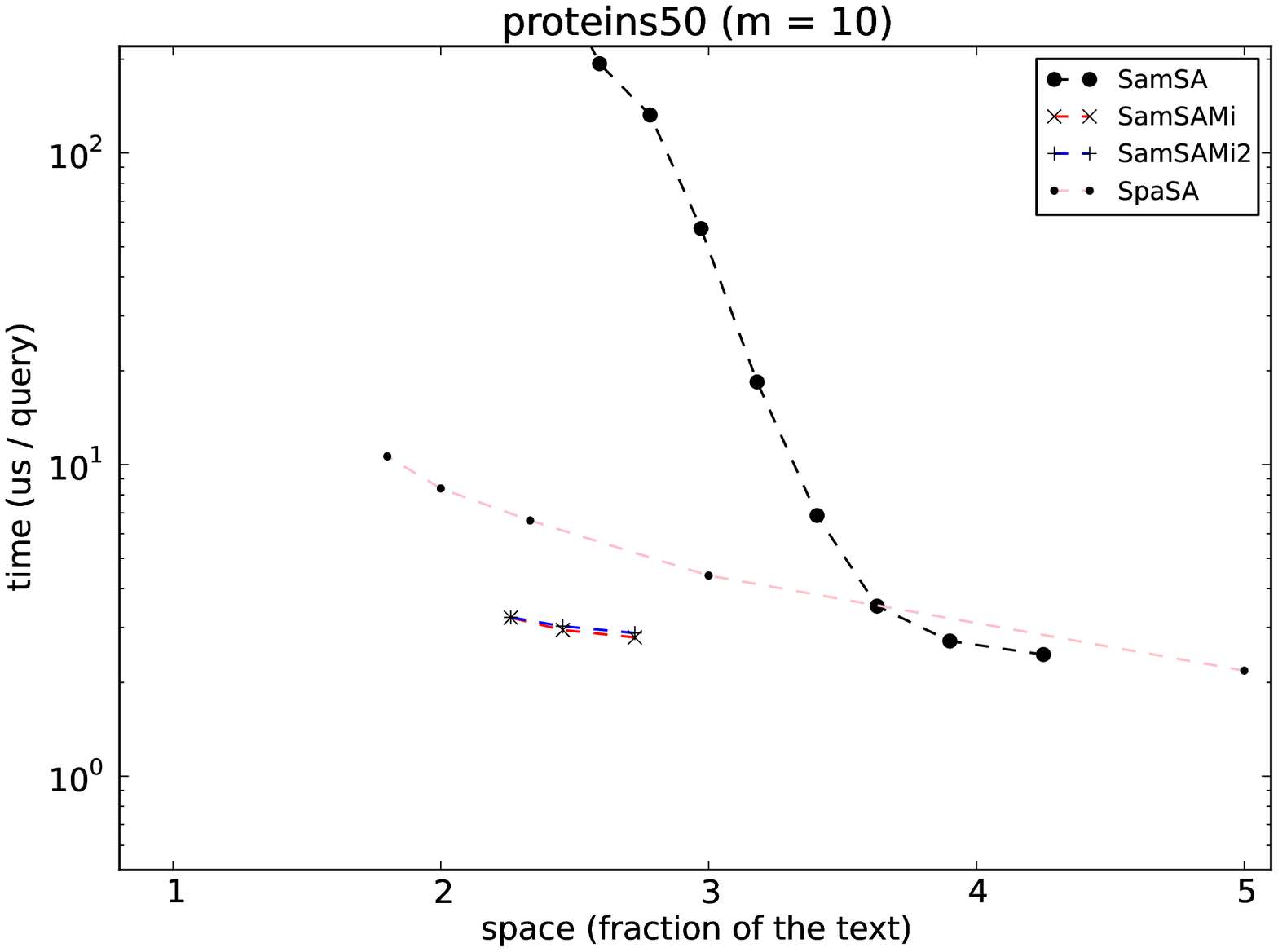}
\includegraphics[width=0.49\textwidth,scale=1.0]{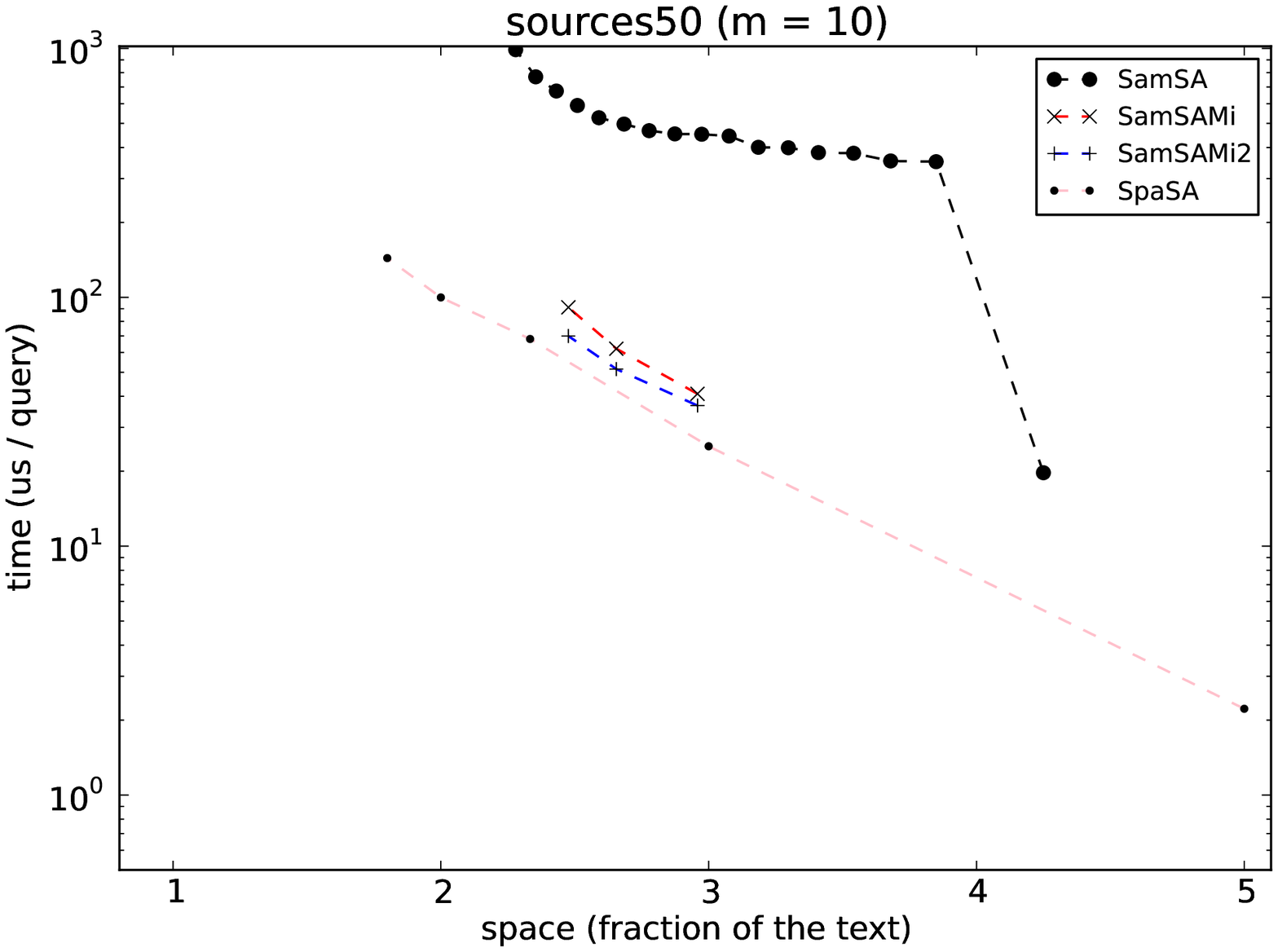}
}
\centerline{
\includegraphics[width=0.49\textwidth,scale=1.0]{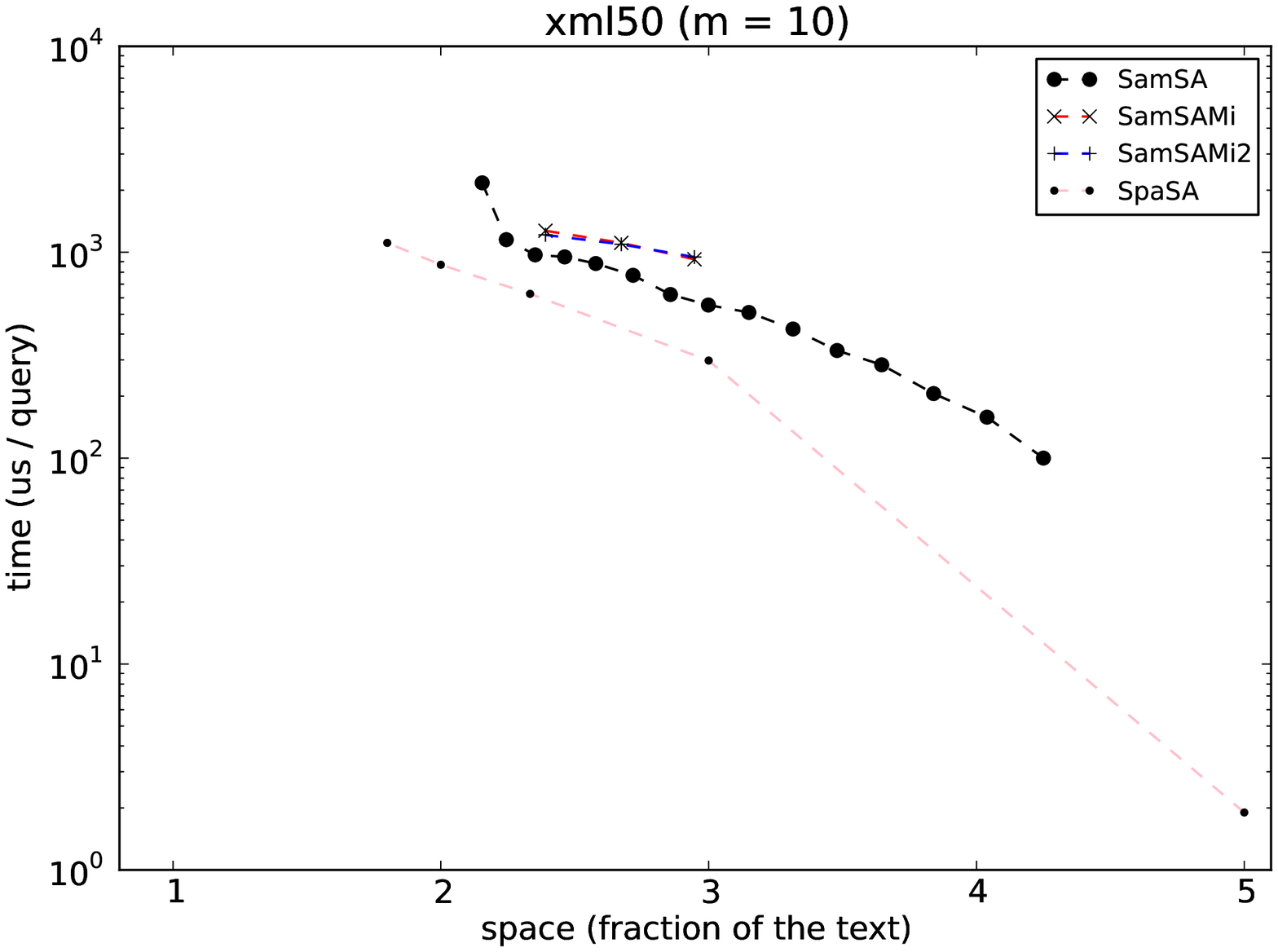}
}
\caption[Results]
{Pattern search time (count query). 
All times are averages over 500K random patterns of length 10.
The patterns were extracted from the respective texts.
Times are given in microseconds.
The index space is a multiple of the text size, including the text.}
\label{fig:times_10}
\end{figure}

\begin{figure}[pt]
\centerline{
\includegraphics[width=0.49\textwidth,scale=1.0]{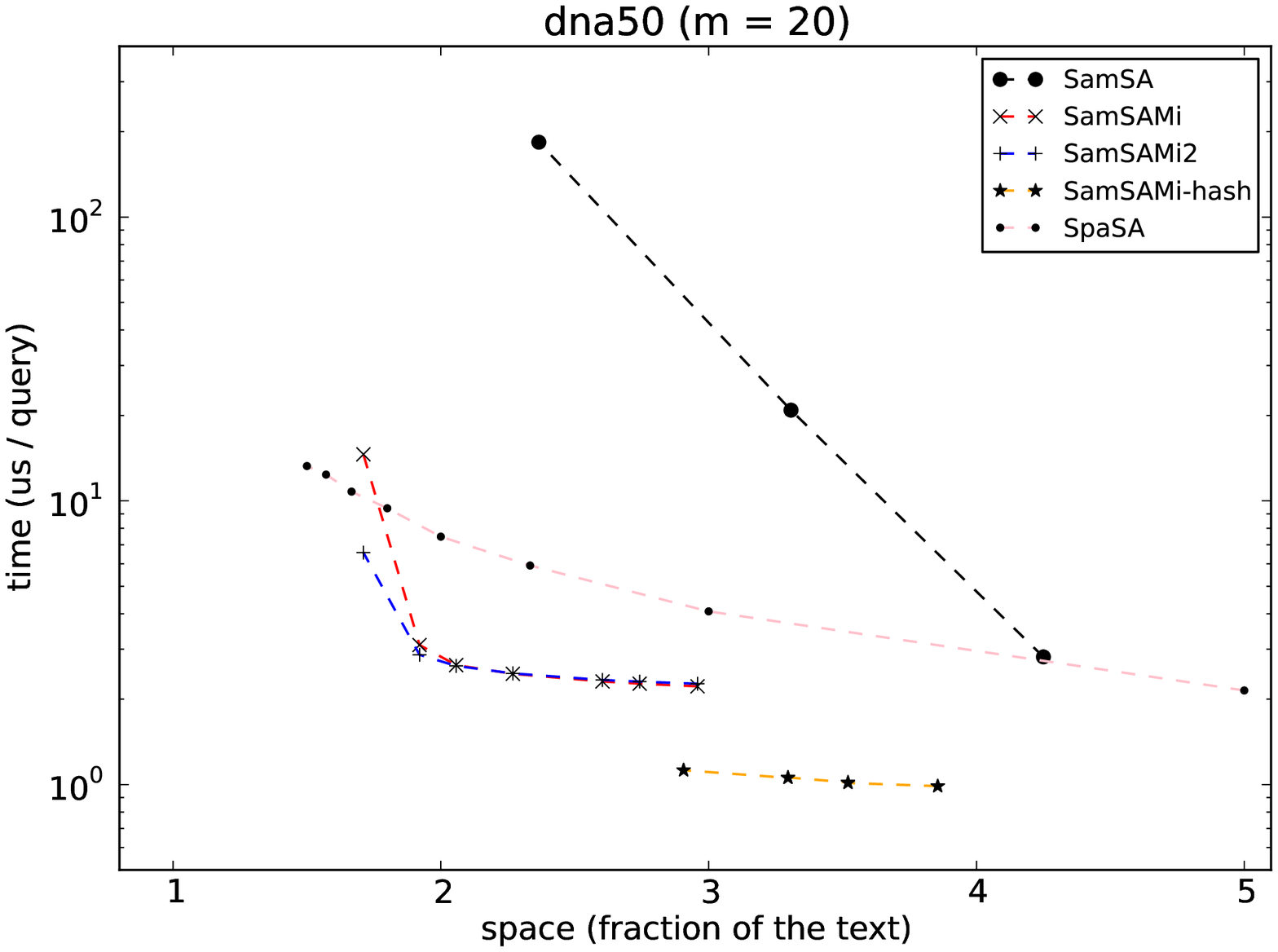}
\includegraphics[width=0.49\textwidth,scale=1.0]{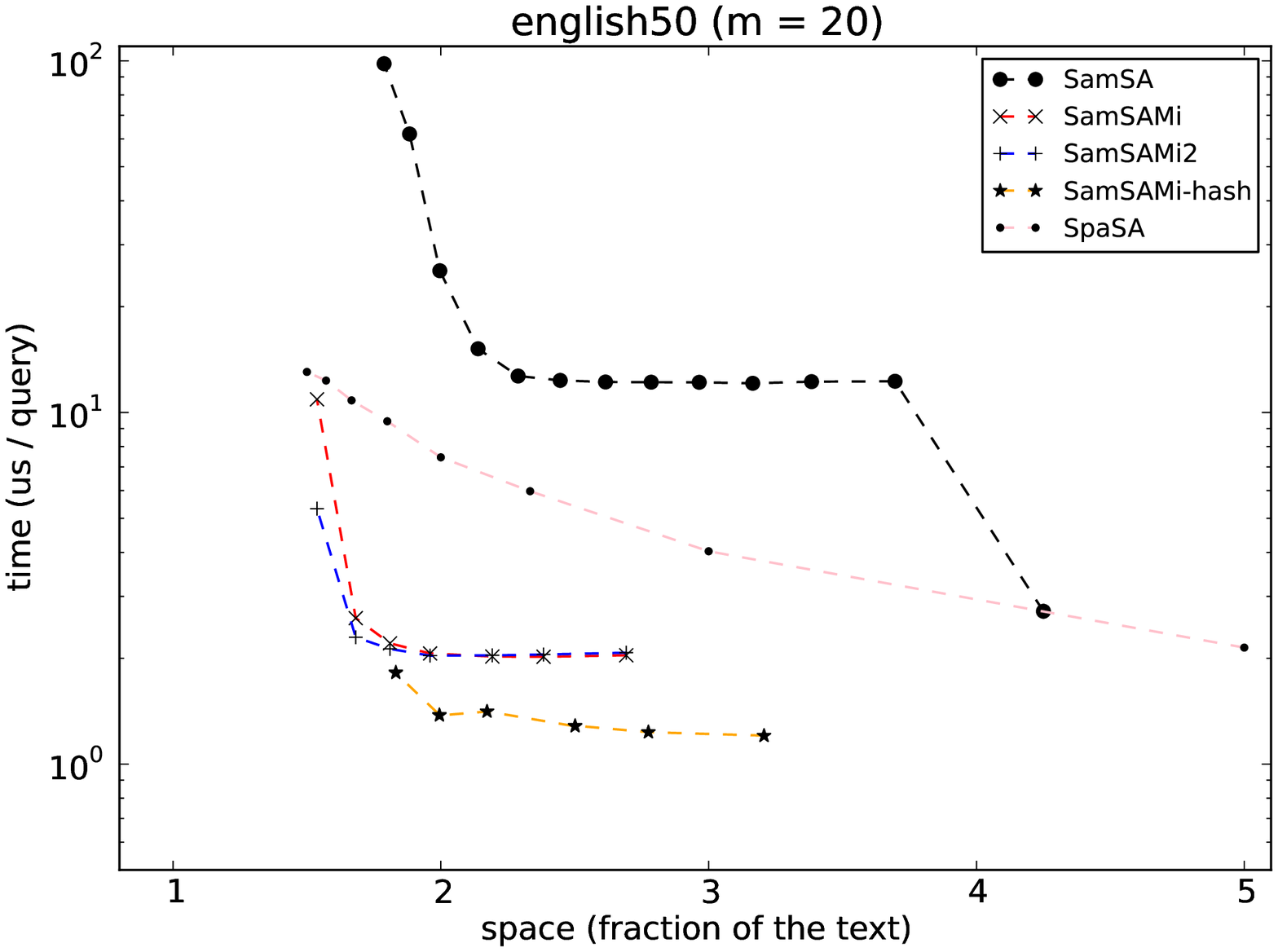}
}
\centerline{
\includegraphics[width=0.49\textwidth,scale=1.0]{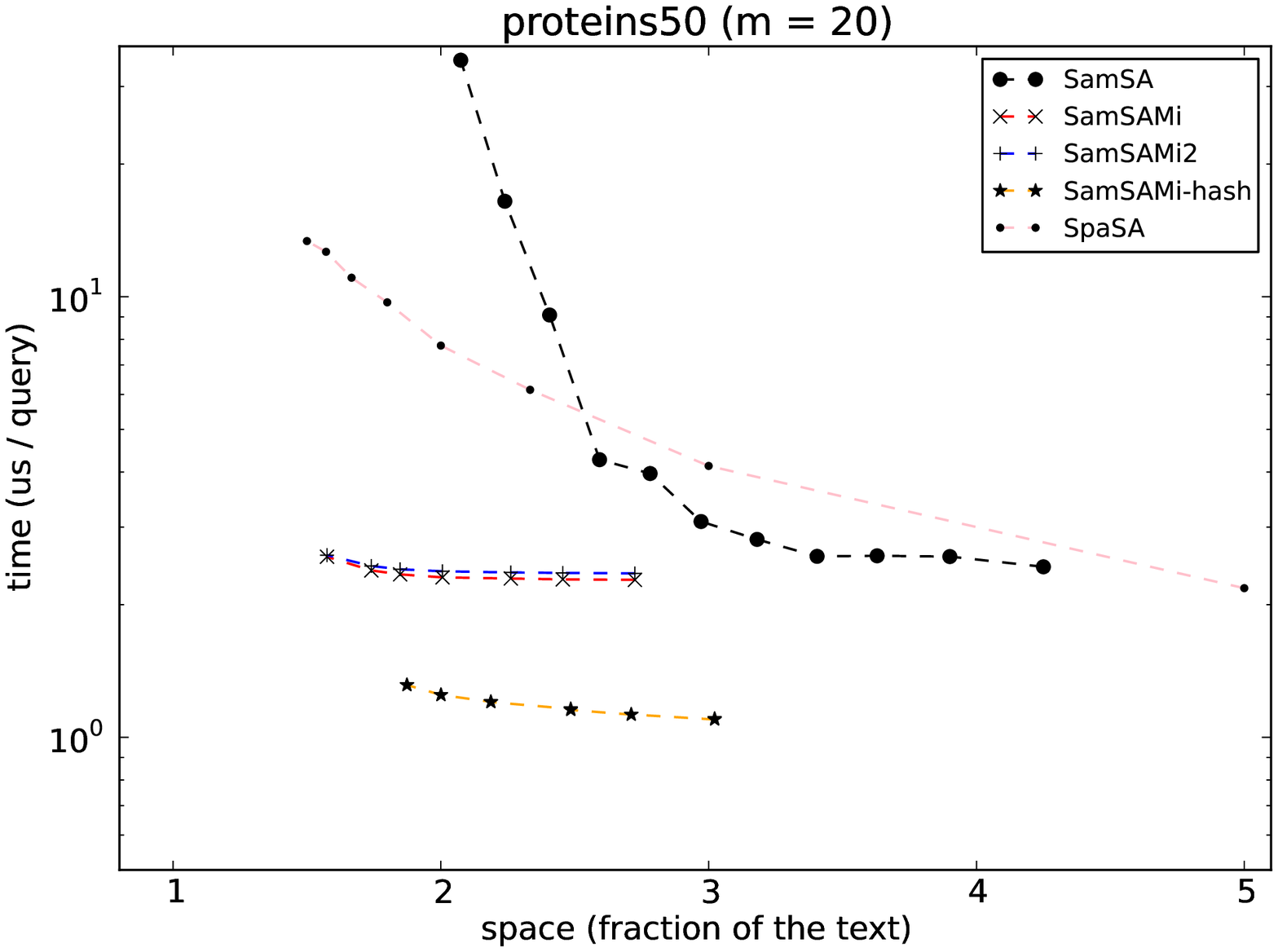}
\includegraphics[width=0.49\textwidth,scale=1.0]{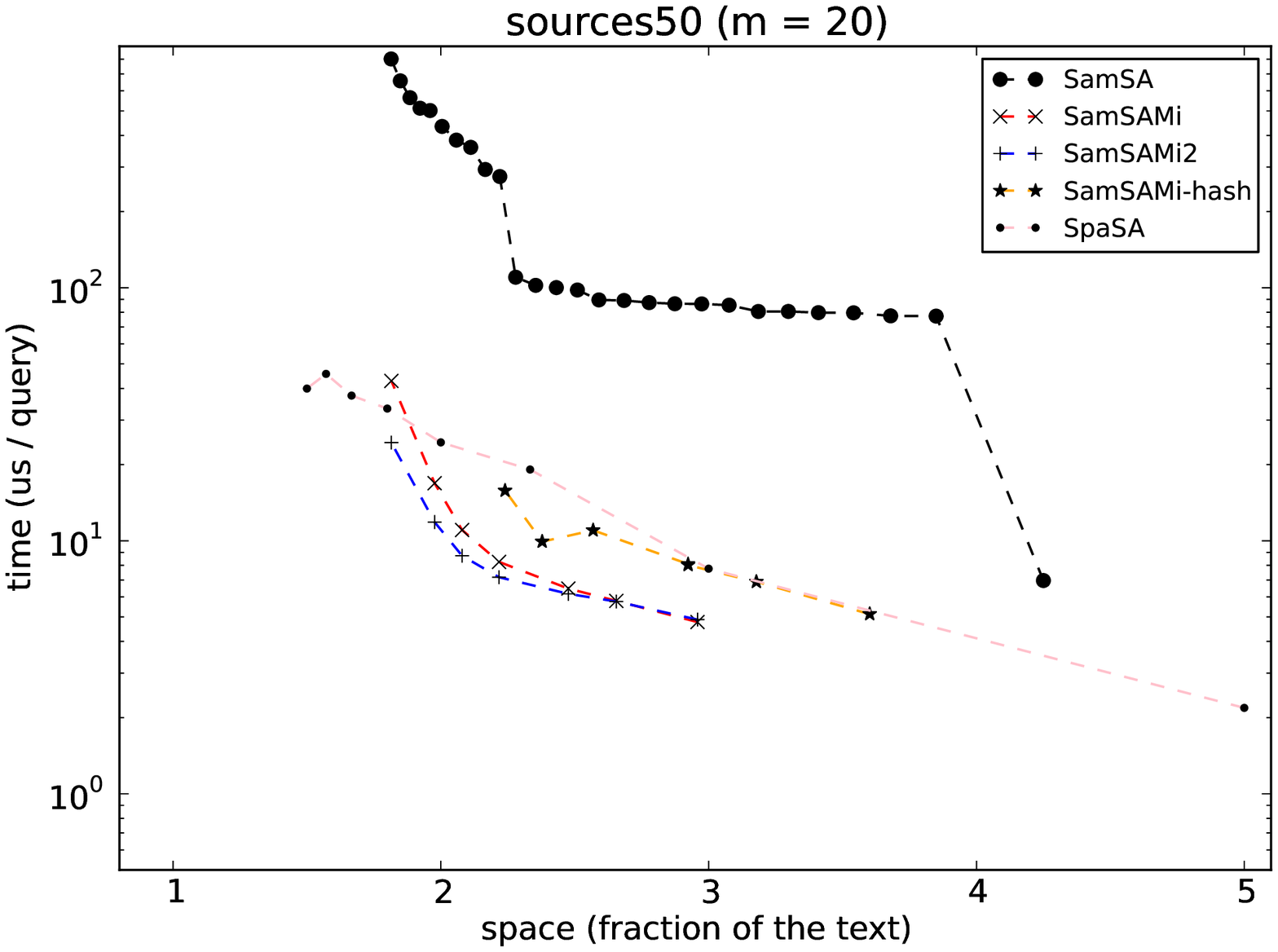}
}
\centerline{
\includegraphics[width=0.49\textwidth,scale=1.0]{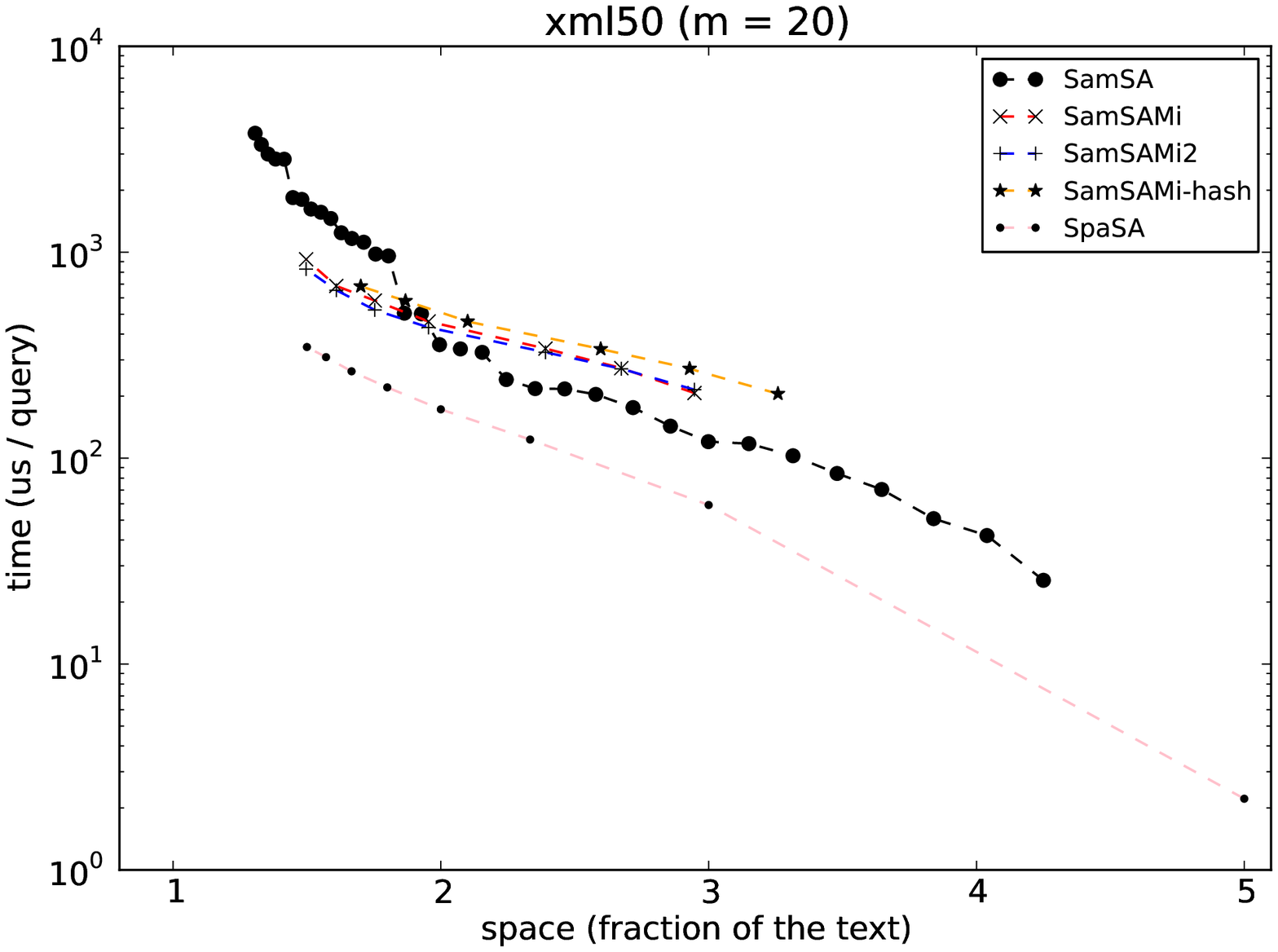}
}
\caption[Results]
{Pattern search time (count query). 
All times are averages over 500K random patterns of length 20.
The patterns were extracted from the respective texts.
Times are given in microseconds.
The index space is a multiple of the text size, including the text.}
\label{fig:times_20}
\end{figure}

\begin{figure}[pt]
\centerline{
\includegraphics[width=0.49\textwidth,scale=1.0]{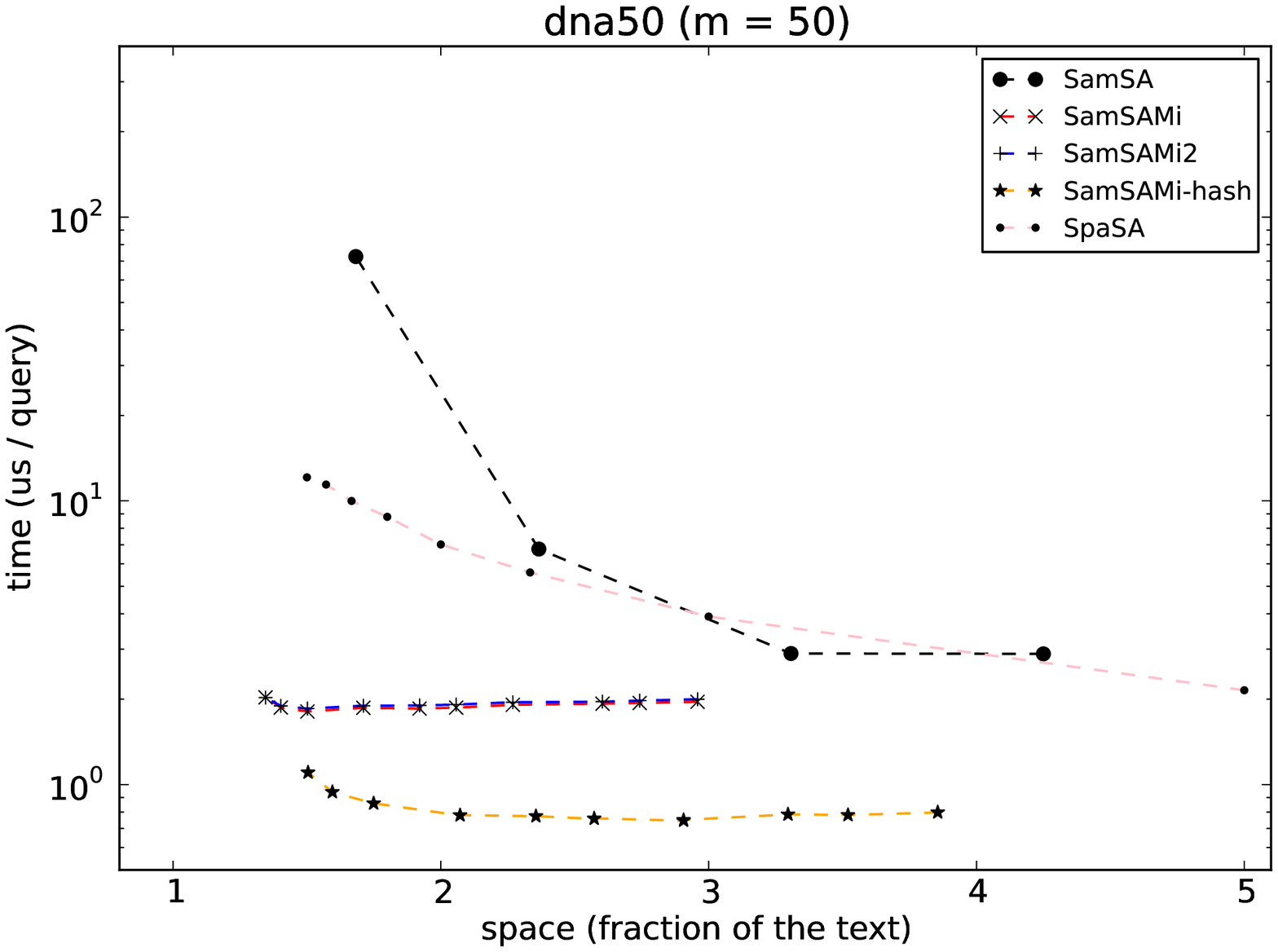}
\includegraphics[width=0.49\textwidth,scale=1.0]{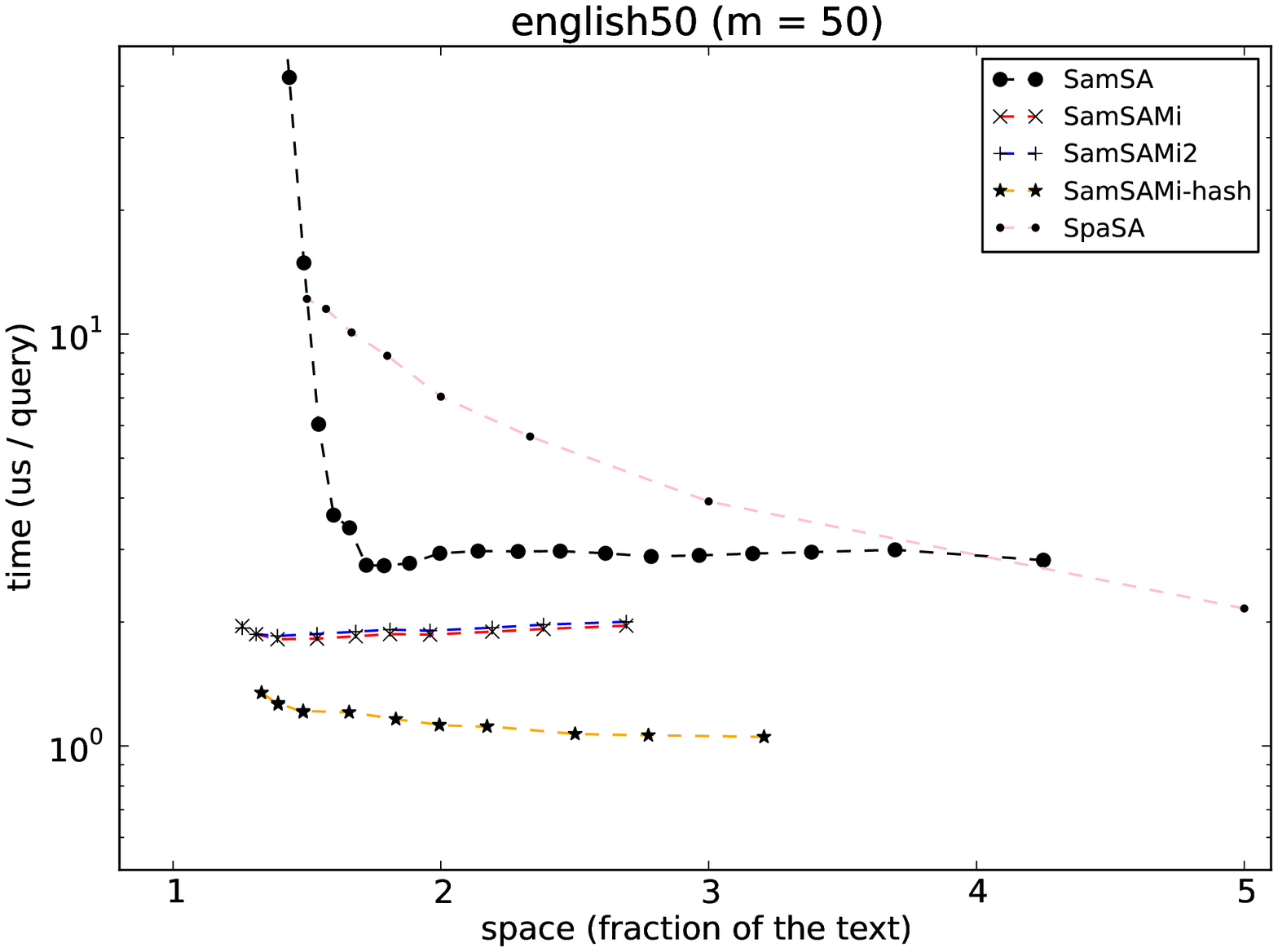}
}
\centerline{
\includegraphics[width=0.49\textwidth,scale=1.0]{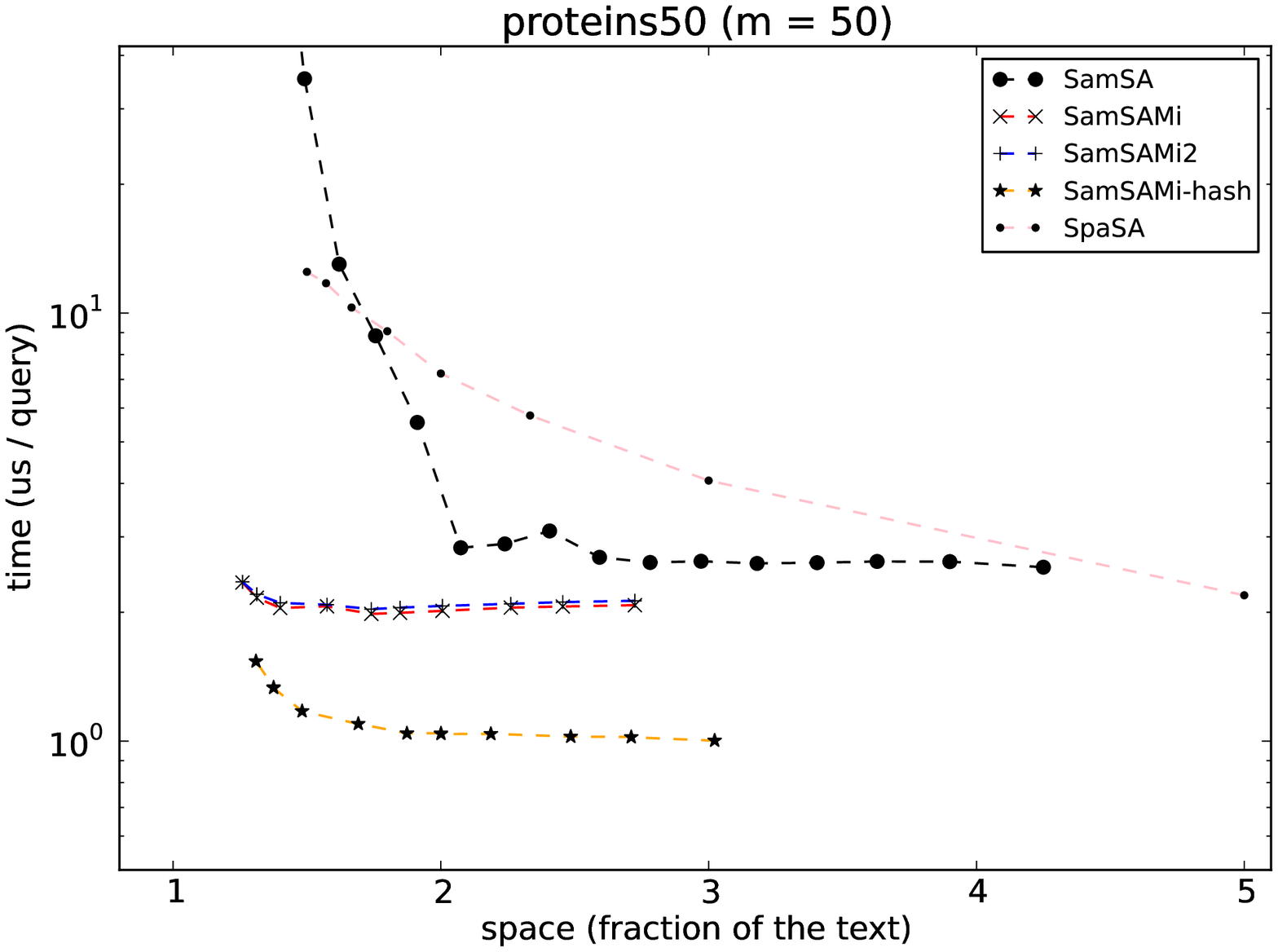}
\includegraphics[width=0.49\textwidth,scale=1.0]{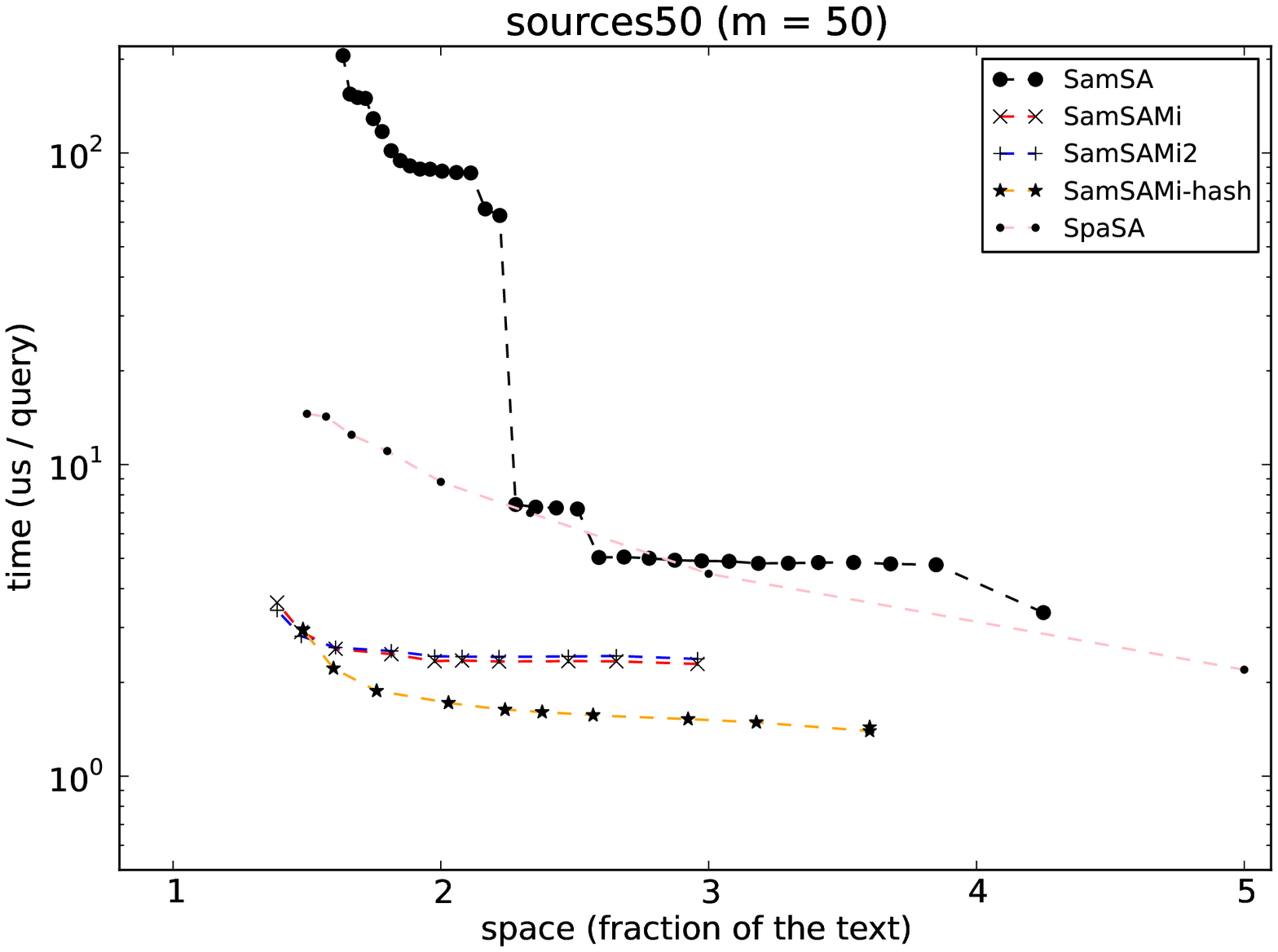}
}
\centerline{
\includegraphics[width=0.49\textwidth,scale=1.0]{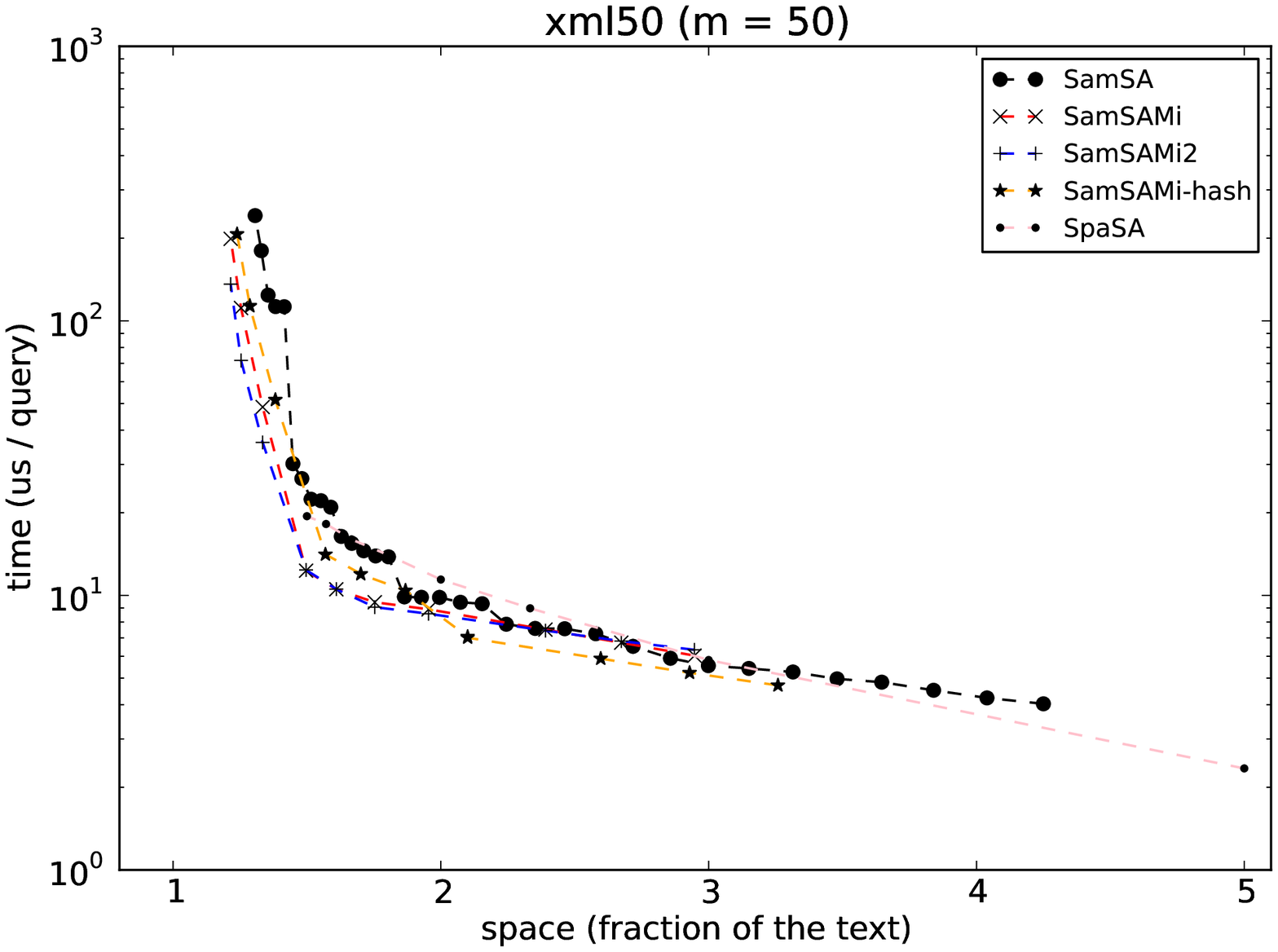}
}
\caption[Results]
{Pattern search time (count query). 
All times are averages over 500K random patterns of length 50.
The patterns were extracted from the respective texts.
Times are given in microseconds.
The index space is a multiple of the text size, including the text.}
\label{fig:times_50}
\end{figure}

\begin{figure}[pt]
\centerline{
\includegraphics[width=0.49\textwidth,scale=1.0]{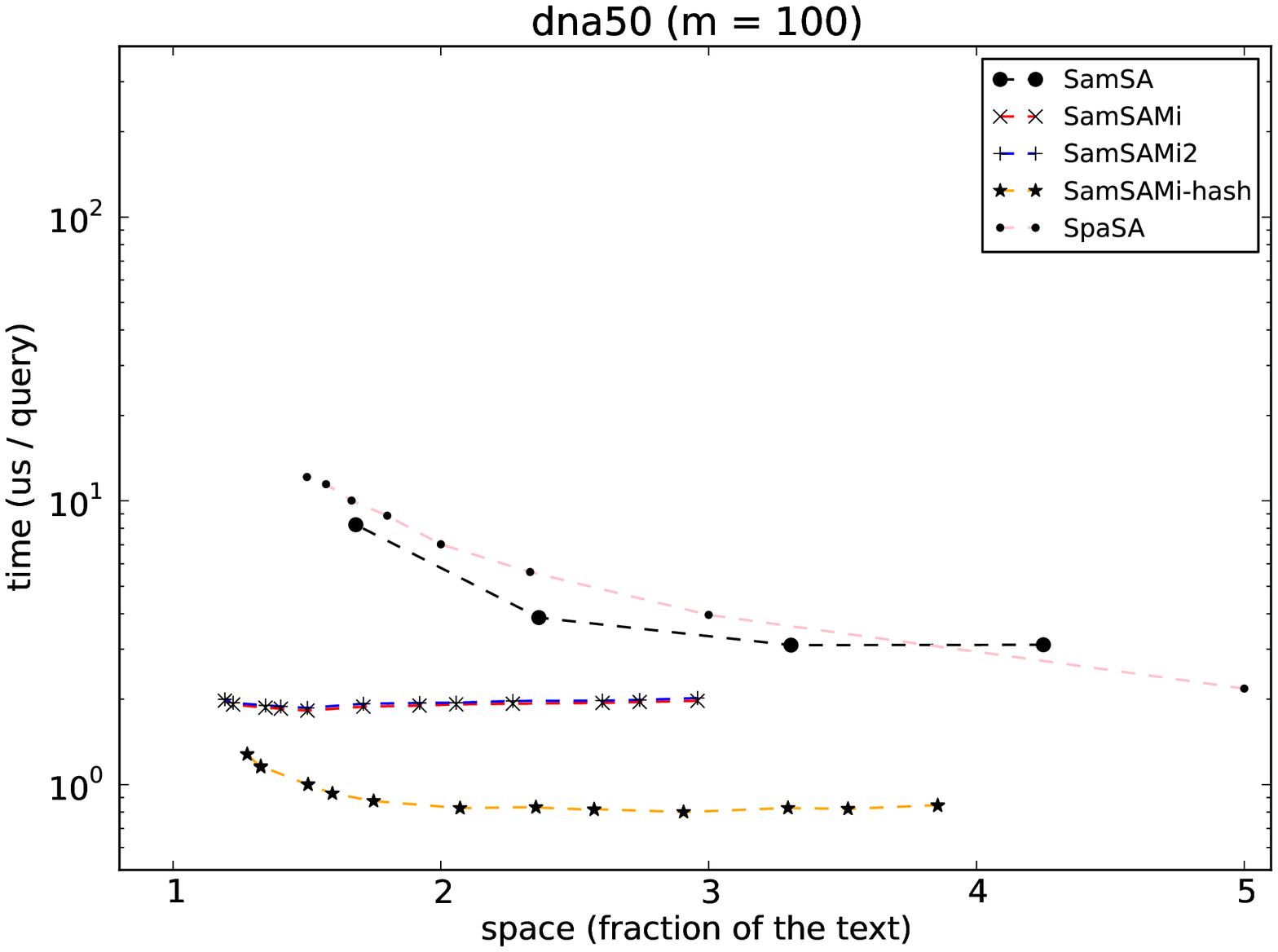}
\includegraphics[width=0.49\textwidth,scale=1.0]{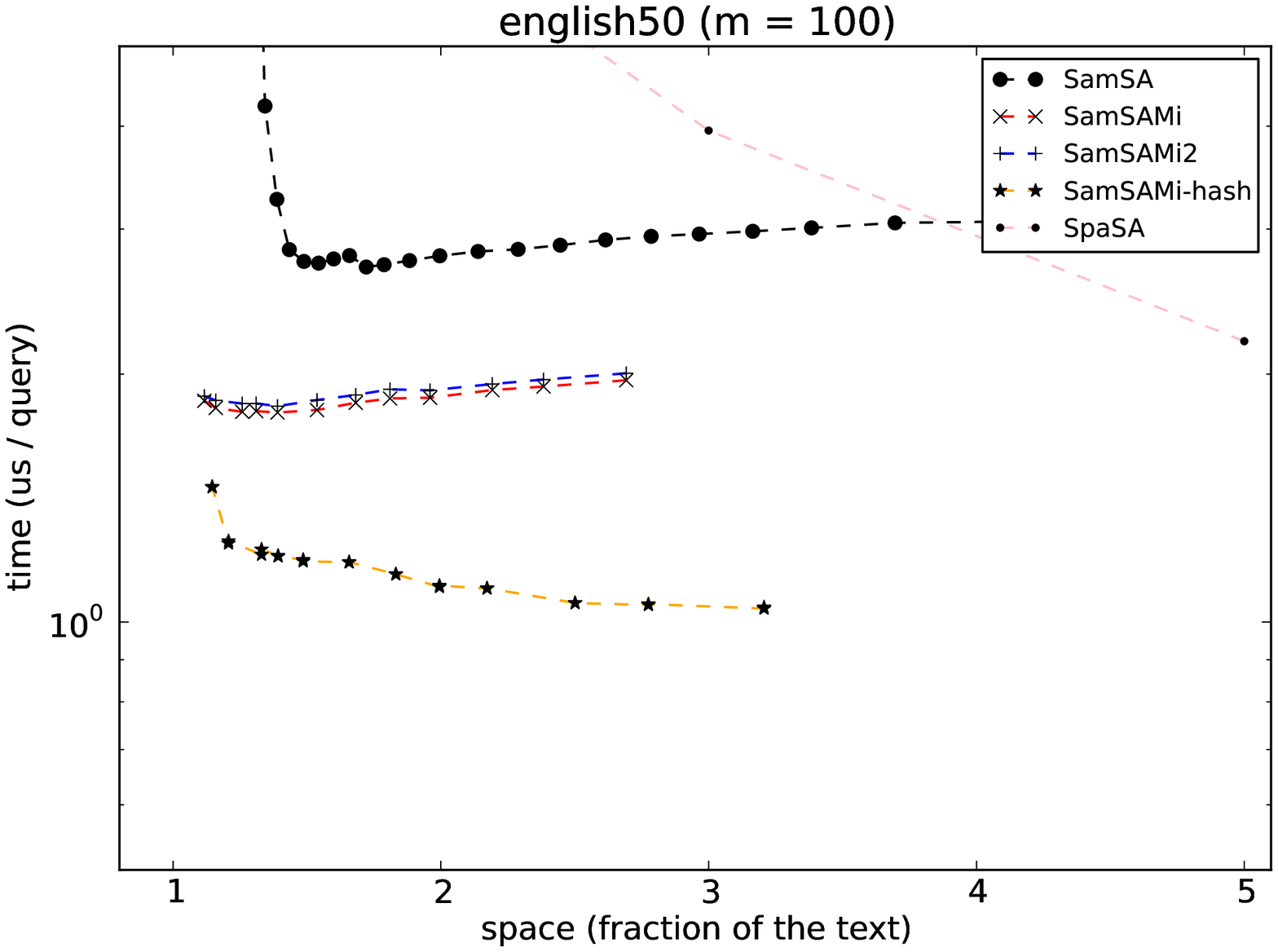}
}
\centerline{
\includegraphics[width=0.49\textwidth,scale=1.0]{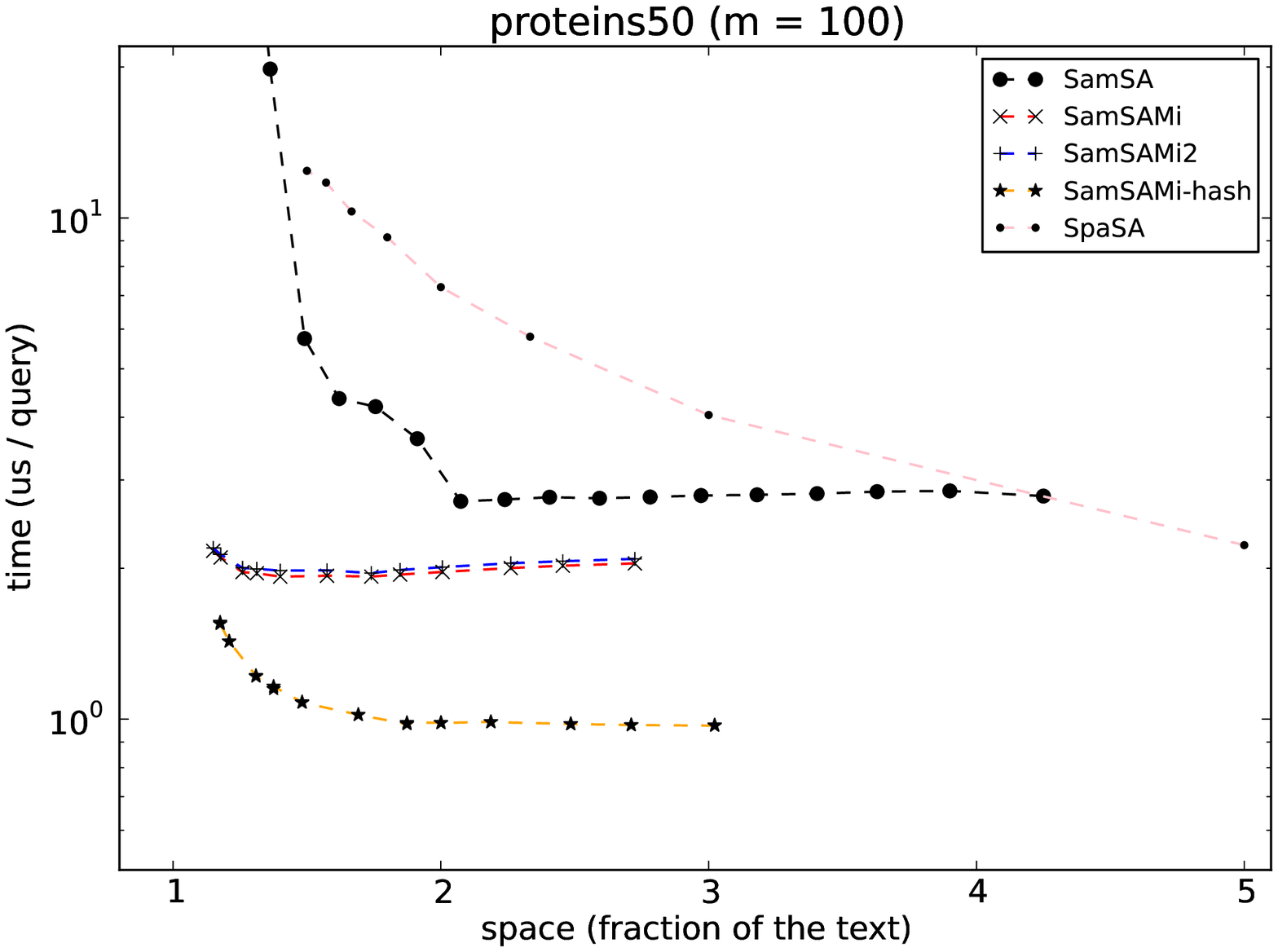}
\includegraphics[width=0.49\textwidth,scale=1.0]{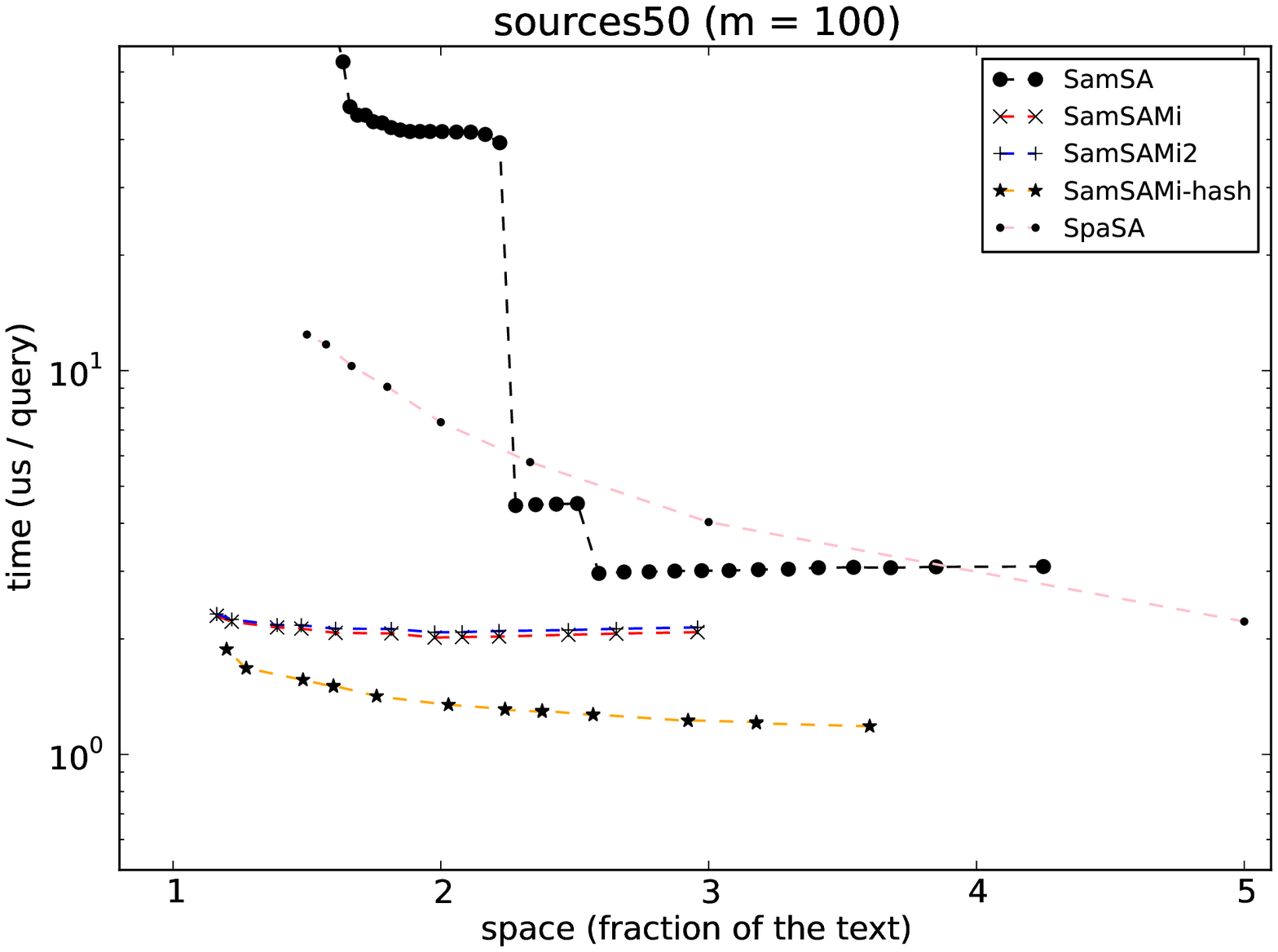}
}
\centerline{
\includegraphics[width=0.49\textwidth,scale=1.0]{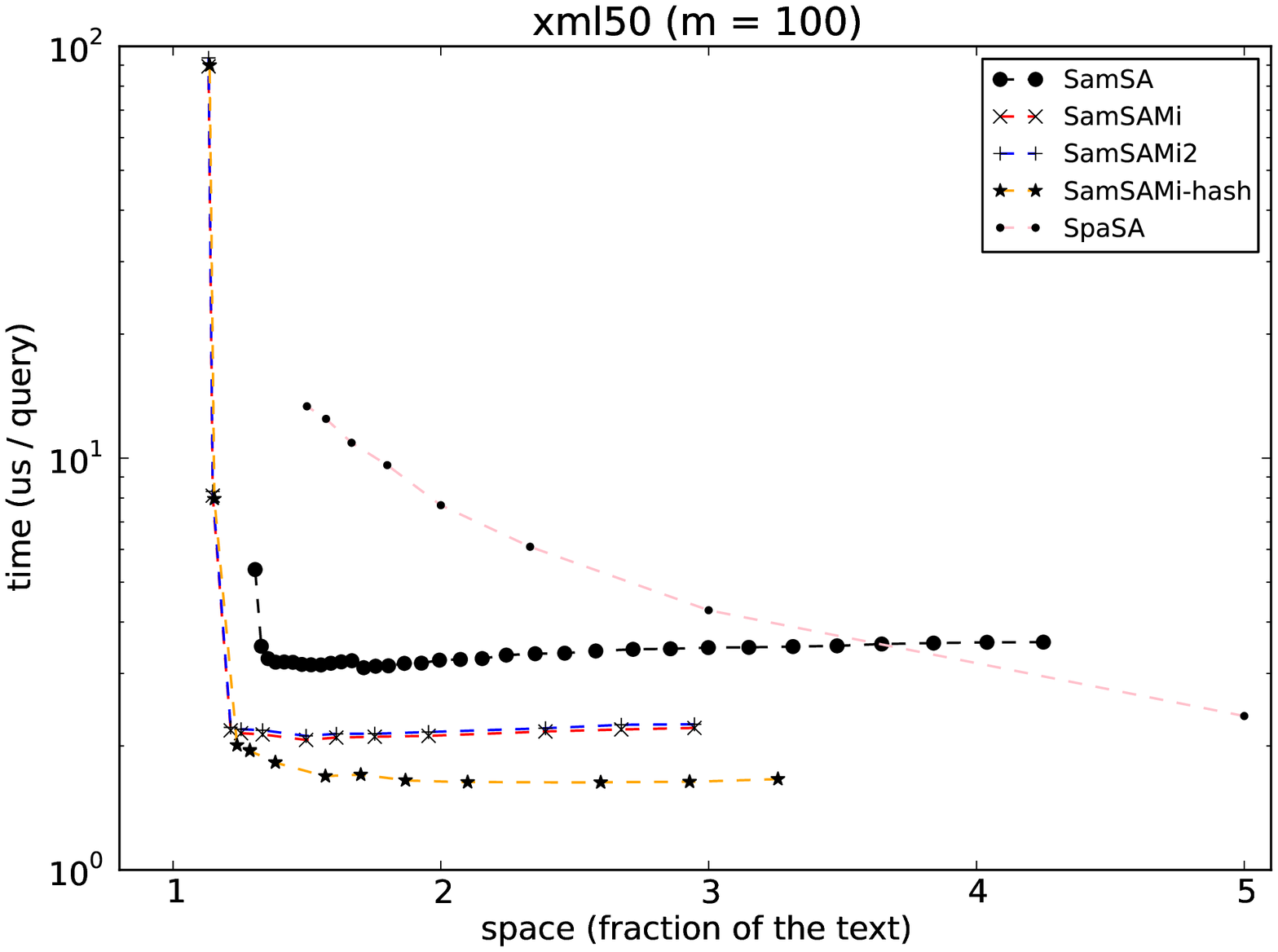}
}
\caption[Results]
{Pattern search time (count query). 
All times are averages over 500K random patterns of length 100.
The patterns were extracted from the respective texts.
Times are given in microseconds.
The index space is a multiple of the text size, including the text.}
\label{fig:times_100}
\end{figure}

Pattern searches were run for $m \in \{10, 20, 50, 100\}$ 
and for each dataset and pattern length 500,000 randomly extracted patterns 
from the text were used.
Figs~\ref{fig:times_10}--\ref{fig:times_100} present average search times 
with respected to varying parameters.
For SpaSA we changed its parameter $k$ from 1 (which corresponds to the 
plain suffix array) to 8.
For SamSAMi we varied $q$ from $\{4, 5, 6, 8, 10, 12, 16, 24, 32, 40, 64, 80\}$ 
setting the most appropriate $p$ (up to 3 or 4) to obtain the smallest index, 
according to the statistics from Table~\ref{table:pq}.
Obviously, $q$ was limited for $m < 100$; up to 6 for $m=10$, up to 16 for $m=20$, 
and up to 40 for $m=50$.

We note that SamSAMi is rather competitive against the sparse suffix array, 
with two exceptions: short patterns ($m=10$) and the XML dataset (for $m=10$ 
and $m=20$).
In most cases, SamSAMi is also competitive against the sampled suffix array, 
especially when aggressive suffix sampling is applied.
(For a honest comparison one should also notice that our implementation 
uses 32-bit suffix indexes while the Claude et al. scheme was tested with 
$\lceil \log_2 n \rceil$ bits per index, which is 26 bits for the used 
datasets.)

Unfortunately, the variant with reduced verifications (\texttt{SamSAMi2}) 
is not significantly faster than the original one, only in rare cases, 
with a large value of the used $q$, the search time can be approximately halved.
SamSAMi-hash, on the other hand, can be an attractive alternative, 
similarly as SA-hash used as a replacement of the plain SA~\cite{GR2014}.

\section{Conclusions and future work}
\label{sec:concl}

We presented a simple suffix sampling scheme making it possible to search 
for patterns effectively.
The resulting data structure, called a sampled suffix array with minimizers 
(SamSAMi), achieves interesting time-space tradeoffs; 
for example, on English50 dataset the search for patterns of length 50 
is still by about 10\% faster than with a plain suffix array 
when only 5.3\% of the suffixes are retained.

Apart from extra experiments, 
several aspects of our ideas require further research.
We mentioned a theoretical solution for building our sampled suffix array 
in small space can be applied, but it is an interesting question if we can 
make use of our parsing properties to obtain $O(n)$ time and $O(n')$ space 
in the worst case.
Such complexities are possible for the suffix array on $n'$ words, 
as shown by Ferragina and Fischer~\cite{FFcpm07}, and their idea can easily  
be used for the sampled SA by Claude et al.~\cite{CNPSTjda10} as noted 
in the cited work.

How to find minimizers efficiently, both in a static sequence 
(i.e., a pattern prefix) and a sliding window, is also of some interest.
Na{\"i}ve implementations result in $O(pq)$ and $O(npq)$ times, respectively, 
but with a heap the latter can be reduced to $O(np\log q)$.
One solution to get rid of the 
factor $q$ 
can be to use the Rabin-Karp 
rolling hash~\cite[Sect.~32.2]{Cormen2009} over the substrings of 
length $p$ and find the minimum hash 
value rather than the lexicographically lowest substring.
Also, a heap may be replaced with a trie storing the $p$-grams. 
Assuming constant-time parent-child navigation over the trie (i.e., also 
a small enough alphabet), we update the trie for one shift of the window 
in $O(p)$ time (as one $p$-gram is removed, one $p$-gram is added, and 
the minimizer is the leftmost string in the trie), which results in $O(np)$ 
overall time.
For finding the minimizer in the pattern prefix we may use a 
rather theoretical option involving linear-time suffix sorting.
To this end, we first remap the pattern prefix alphabet to (at most) 
$\{0, 1, \ldots, q-1\}$, in $O(q\log q)$ time, using a balanced BST.
Next we sort the pattern prefix suffixes (using, e.g., the algorithm from~\cite{NZC11})
and finally scan over the sorted list of suffixes until the 
first suffix of length at least $p$ is found.
The total time is $O(q\log q)$.


\section*{Acknowledgement}
We thank Kimmo Fredriksson for helpful comments and Francisco Claude 
for sharing with us the sampled suffix array sources.

The work was supported by the Polish Ministry of Science and Higher Education under the project DEC-2013/09/B/ST6/03117 (both authors).

\bibliographystyle{abbrv}
\bibliography{misasa}

\begin{thebibliography}{10}

\bibitem{ABR00}
S.~Alstrup, G.~S. Brodal, and T.~Rauhe.
\newblock Pattern matching in dynamic texts.
\newblock In {\em Proceedings of the 11th Annual Symposium on Discrete
  Algorithms (SODA)}, pages 819--828. Society for Industrial and Applied
  Mathematics, 2000.

\bibitem{BFNP2007}
N.~Brisaboa, A.~Fari{\~n}a, G.~Navarro, and J.~Param{\'a}.
\newblock Lightweight natural language text compression.
\newblock {\em Information Retrieval}, 10:1--33, 2007.

\bibitem{CLJSM2014}
R.~Chikhi, A.~Limasset, S.~Jackman, J.~Simpson, and P.~Medvedev.
\newblock On the representation of de {B}ruijn graphs.
\newblock {\em arXiv preprint arXiv:1401.5383}, 2014.

\bibitem{CNPSTjda10}
F.~Claude, G.~Navarro, H.~Peltola, L.~Salmela, and J.~Tarhio.
\newblock String matching with alphabet sampling.
\newblock {\em Journal of Discrete Algorithms (JDA)}, 11:37--50, 2012.

\bibitem{Cormen2009}
T.~H. Cormen, C.~E. Leiserson, R.~L. Rivest, and C.~Stein.
\newblock {\em Introduction to Algorithms (3. ed.)}.
\newblock MIT Press, 2009.

\bibitem{CLGLPR00}
P.~Crescenzi, A.~D. Lungo, R.~Grossi, E.~Lodi, L.~Pagli, and G.~Rossi.
\newblock Text sparsification via local maxima.
\newblock In {\em Proceedings of the 20th Conference on the Foundations of
  Software Technology and Theoretical Computer Science(FSTTCS)}, volume 1974 of
  {\em LNCS}, pages 290--301. Springer, 2000.

\bibitem{CLGLPR03}
P.~Crescenzi, A.~D. Lungo, R.~Grossi, E.~Lodi, L.~Pagli, and G.~Rossi.
\newblock Text sparsification via local maxima.
\newblock {\em Theoretical Computer Science}, 1--3(304):341--364, 2003.

\bibitem{FFcpm07}
P.~Ferragina and J.~Fischer.
\newblock Suffix arrays on words.
\newblock In {\em CPM}, volume 4580 of {\em LNCS}, pages 328--339.
  Springer--Verlag, 2007.

\bibitem{FGNV2008}
P.~Ferragina, R.~Gonz{\'a}lez, G.~Navarro, and R.~Venturini.
\newblock Compressed text indexes: From theory to practice.
\newblock {\em ACM Journal of Experimental Algorithmics (JEA)}, 13:article 12,
  2009.
\newblock 30 pages.

\bibitem{FG2006}
K.~Fredriksson and S.~Grabowski.
\newblock A general compression algorithm that supports fast searching.
\newblock {\em Information Processing Letters}, 100(6):226--232, 2006.

\bibitem{GP2013}
S.~Gog and M.~Petri.
\newblock Optimized succinct data structures for massive data.
\newblock {\em Software--Practice and Experience}, 2013.
\newblock DOI: 10.1002/spe.2198.

\bibitem{GR2014}
S.~Grabowski and M.~Raniszewski.
\newblock Two simple full-text indexes based on the suffix array.
\newblock {\em CoRR}, abs/1405.5919, 2014.

\bibitem{HT1971}
T.~C. Hu and A.~C. Tucker.
\newblock Optimal computer search trees and variable-length alphabetical codes.
\newblock {\em SIAM Journal on Applied Mathematics}, 21(4):514--532, 1971.

\bibitem{IKK2014}
T.~I, J.~K{\"a}rkk{\"a}inen, and D.~Kempa.
\newblock Faster sparse suffix sorting.
\newblock In {\em STACS}, volume~25 of {\em LIPIcs}, pages 386--396. Schloss
  Dagstuhl - Leibniz-Zentrum fuer Informatik, 2014.

\bibitem{DBLP:conf/cocoon/KarkkainenU96}
J.~K{\"a}rkk{\"a}inen and E.~Ukkonen.
\newblock Sparse suffix trees.
\newblock In {\em COCOON}, volume 1090 of {\em LNCS}, pages 219--230, 1996.

\bibitem{LKHYYS2013}
Y.~Li, P.~Kamousi, F.~Han, S.~Yang, X.~Yan, and S.~Suri.
\newblock Memory efficient minimum substring partitioning.
\newblock In {\em Proceedings of the 39th International Conference on Very
  Large Data Bases}, pages 169--180. VLDB Endowment, 2013.

\bibitem{MSU97}
K.~Mehlhorn, R.~Sundar, and C.~Uhrig.
\newblock Maintaining dynamic sequences under equality tests in polylogarithmic
  time.
\newblock {\em Algorithmica}, 17(2):183--198, 1997.

\bibitem{MFC2012}
N.~S. Movahedi, E.~Forouzmand, and H.~Chitsaz.
\newblock De novo co-assembly of bacterial genomes from multiple single cells.
\newblock In {\em BIBM}, pages 1--5, 2012.

\bibitem{NM2007}
G.~Navarro and V.~M{\"a}kinen.
\newblock Compressed full-text indexes.
\newblock {\em ACM Comput. Surv.}, 39(1):article 2, 2007.

\bibitem{NZC11}
G.~Nong, S.~Zhang, and W.~H. Chan.
\newblock Two efficient algorithms for linear time suffix array construction.
\newblock {\em IEEE Transactions on Computers}, 60(10):1471--84, 2011.

\bibitem{PST2006}
S.~J. Puglisi, W.~F. Smyth, and A.~Turpin.
\newblock Inverted files versus suffix arrays for locating patterns in primary
  memory.
\newblock In {\em SPIRE}, volume 4209 of {\em LNCS}, pages 122--133, 2006.

\bibitem{RHHMY2004}
M.~Roberts, W.~Hayes, B.~R. Hunt, S.~M. Mount, and J.~A. Yorke.
\newblock Reducing storage requirements for biological sequence comparison.
\newblock {\em Bioinformatics}, 20(18):3363--3369, 2004.

\bibitem{SV94}
S.~C. Sahinalp and U.~Vishkin.
\newblock Symmetry breaking for suffix tree construction.
\newblock In {\em Proceedings of the 26th Annual ACM Symposium on Theory of
  Computing (STOC)}, pages 300--309. ACM, 1994.

\bibitem{WS2014}
D.~E. Wood and S.~L. Salzberg.
\newblock Kraken: ultrafast metagenomic sequence classification using exact
  alignments.
\newblock {\em Genome Biology}, 15(3):R46, 2014.

\end{thebibliography}

\end{document}